# Observation of a possible diluted ferromagnetism above room temperature in cobalt-substituted $LaTa(O,N)_{3-\delta}$

Cora Bubeck,[1,2,*] Eberhard Goering,[3,*,†] Robert Lawitzki,[2] Kathrin Küster,[4] Wilfried Sigle,[4] Marc Widenmeyer,[1] Ulrich Starke,[4] Clemens Ritter,[5] Gabriel J. Cuello,[5] Peter Nagel,[6,7] Michael Merz,[6,7] Stefan Schuppler,[6,7] Gisela Schütz[3] & Anke Weidenkaff[1,8,‡]

[1]*Technical University of Darmstadt, Department of Materials and Earth Sciences, Materials and Resources, Alarich-Weiss-Straße 2, 64287 Darmstadt, Germany*

[2]*University of Stuttgart, Institute for Materials Science, Heisenbergstraße 3, 70569 Stuttgart, Germany*

[3]*Max Planck Institute for Intelligent Systems, Modern Magnetic Systems, Heisenbergstraße 3, 70569 Stuttgart, Germany*

[4]*Max Planck Institute for Solid State Research, Heisenbergstraße 1, 70569 Stuttgart, Germany*

[5]*Institut Laue Langevin, 71 Avenue des Martyrs, 38042 Grenoble Cédex 9, France*

[6]*Karlsruhe Institute of Technology (KIT), Institute for Quantum Materials and Technologies (IQMT), 76021 Karlsruhe, Germany*

[7]*Karlsruhe Nano Micro Facility (KNMFi), Karlsruhe Institute for Technology (KIT), 76344 Eggenstein-Leopoldshafen, Germany*

[8]*Fraunhofer Research Institution for Materials Recycling and Resource Strategies IWKS, Brentanostr. 2a, 63755 Alzenau, Germany*

Since 2000, the intensive effort in materials research to develop a diluted magnetic semiconductor exhibiting high-temperature (HT) ferromagnetism above room temperature was not successful. Here, the possible first bulk diluted HT-ferromagnetic non-metallic materials, based on the perovskite-type oxynitrides $LaTa_{1-x}Co_x(O,N)_{3-\delta}$ ($x$ = 0.01, 0.03, 0.05) are realized. The Curie temperature of the synthesized powders exceeds 600 K and the sample magnetizations are large enough to be directly attracted by permanent magnets. Cobalt clusters as a possible source for the observed HT-ferromagnetism can be excluded, since all applied characterization methods verify phase purity. Applied conventional and element-specific magnetometry imply ferromagnetic intermediate spin (IS) $Co^{3+}$ which is included in a ferromagnetic host matrix. This indicates a complex magnetic interplay between the existing crystal structure, the observed anionic vacancies, and the introduced cobalt ions. These results lay the foundation for the experimental investigation and design of further diluted HT-ferromagnetic semiconductors.

## I. INTRODUCTION

Diluted magnetic semiconductors (DMS) such as ferromagnetic *p*-type $Ga_{1-x}Mn_xAs$ are very promising for applications in spintronics [1–4]. However, until now ferromagnetism at room temperature for DMS has not been observed. This desired property would open up big advances in developing multifunctional ferromagnetic devices for spintronics. In the year 2000 Dietl and co-workers predicted the possibility to obtain high-temperature (HT) ferromagnetism above room temperature *via* 3*d* transition metal doping in semiconductors and insulators (*e.g.* in ZnO or GaN) [1,2]. Dietl *et al.* expected that *p*-type materials containing a critical concentration of holes and magnetic ions leading to DMS should exhibit an even higher Curie temperature ($T_C$) than room temperature [1].

The stated prediction is very counterintuitive, because it is well-known that a strong prerequisite for room temperature ferromagnetism is a strong exchange interaction. This is not expected for large ionic distances between magnetic ions and/or holes existing in DMS [5]. Therefore – and despite many investigations during the last two decades – this is one of the most controversial research topics in materials science and condensed-matter physics [2,6].

Until now, it has been observed that homogeneously dissolved 3*d* transition metal ions in a non-magnetic semiconductor showed paramagnetism [7]. In other cases, secondary phases – such as simple metallic transition metal clusters – contributed to the ferromagnetic-like behavior [6,8]. In this context, new interesting magnetic phenomena such as the "$d^0$–magnetism" were found [9–11]. Even materials which were not doped with transition metals, such as pristine ZnO, revealed ferromagnetism [12]. The origin of this unexpected ferromagnetism is attributed to defect states, which are predominantly located at grain boundary sites [12–15]. Several other attempts to obtain HT-ferromagnetism above room temperature by 3*d* transition metal ion doping in non-magnetic semiconductors failed. [2] In particular, the attempt to realize

*These authors contributed equally to this work.

†goering@is.mpg.de

‡anke.weidenkaff@mr.tu-darmstadt.de

HT-ferromagnetic thin films ended up in measuring magnetometer artefacts, contaminations, and ferromagnetic secondary phases [6]. Until now, the observed ferromagnetism in DMS was far below room temperature. Even for $Ga_{1-x}Mn_xAs$ or $Ge_{1-x}Mn_xTe$ the DMS ferromagnetic behavior is only observed below 200 K [2,16], therefore, a room temperature ferromagnetic DMS has not been realized.

To tailor many of the magnetic properties (*e.g.* magnetization, magnetocrystalline anisotropy, etc.) the doping by ions or substitution of ions in a materials' matrix is a powerful tool. Perovskite-type oxynitrides $AB(O,N)_3$ are normally considered to be suitable for visible light-driven applications or as cadmium-free inorganic pigments [17–20]. This is because they exhibit an extraordinary flexibility in *A*- and *B*-site substitution, with which the physical properties can be tuned [17,21]. We showed previously that $LaTa(O,N)_3$ has a clear non-magnetic semiconducting behavior with very small magnetic moments and diamagnetism at room temperature [17]. Therefore, it seems to be a promising non-magnetic matrix material for *B*-site substitution with tiny amounts of magnetic ions such as $Co^{2+}$.

Here, we show the realization of a HT-ferromagnetic bulk DMS by applying Co-substitution in $LaTa(O,N)_3$. We synthesized red perovskite-type oxynitrides $LaTa_{1-x}Co_x(O,N)_{3-\delta}$ (LTCON) with three different Co ion concentrations, namely 0.2 at% ($x = 0.01$), 0.6 at% ($x = 0.03$), and 1 at% ($x = 0.05$), which all exhibit ferromagnetism far above room temperature. By chemical analysis, advanced nanostructural characterization (*e.g.* state-of-the-art high-resolution transmission electron microscopy (HR-TEM)), and by synthesizing a reference sample containing a non-stoichiometric higher Co ion concentration (hereafter called Co-rich), we can rule out metallic (elemental) Co and Co-rich phases as the source for the observed HT-ferromagnetism. SQUID-based magnetometry and high-quality transmission (TR) mode X-ray magnetic circular dichroism (XMCD) proved significant non-metallic Co-related ferromagnetism and intermediate spin (IS) $Co^{3+}$ down to 0.2 at% Co ions. The obtained single-phase HT-ferromagnetic LTCON powders have an optical bandgap $E_G$ between 1.7 eV and 1.9 eV and can be described as DMS because of the tiny Co ion concentrations and anionic vacancies inside.

## II. SINGLE-PHASE PEROVSKIE-TYPE LTCON – SYNTHESIZABLE OR NOT?

The main question in many studies [2] is the solubility limit of ferromagnetic ions such as $Co^{z+}$, $Fe^{z+}$, or $Ni^{z+}$ in the materials matrix. This is a crucial point, because if the majority of the Co ions are solved in the material the total magnetization (HT-ferromagnetism) cannot be related to the small remaining amounts of Co-rich secondary phases. Normally, materials containing such ions – especially oxynitrides – are difficult to synthesize [22]. Hence, an evaluation of the possibility for synthesis of LTCON by calculating the structural stability according to Li *et al.* [23] was done. According to the calculations the determined tolerance factor *t* [23] of LTCON is $0.934(5) \leq t \leq 0.936(9)$. Therefore, the samples should be synthesizable in the perovskite structure and should be single-phase. Additionally, Ta ions and Co ions exhibit similar effective ionic radii [24] making them suitable for a substitution with each other. Hence, by a careful evaluation of the synthesis parameters, single-phase LTCON samples with all three different Co ion concentrations should be possible.

Therefore, we used an appropriate synthesis protocol to produce LTCON: first, the bluish oxide precursors $LaTa_{1-x}Co_xO_{4-\delta}$ (LTCO) were prepared *via* a sol-gel-related method (Pechini method) (Fig. 1). Afterwards, the LTCO powders were ammonolyzed (heating under flowing $NH_3$ gas) in order to obtain LTCON. A more detailed chemical analysis of the LTCO precursors and formation of LTCON is described in the Supplemental Material Sec. I–II, Fig. S1–S12 and Tabs. S1–S4. After the synthesis of LTCON, we have chosen a series of complementary investigation techniques in order to give a complete picture of the single-phase nature of the materials and their ferromagnetic behavior. This is described in the following sections.

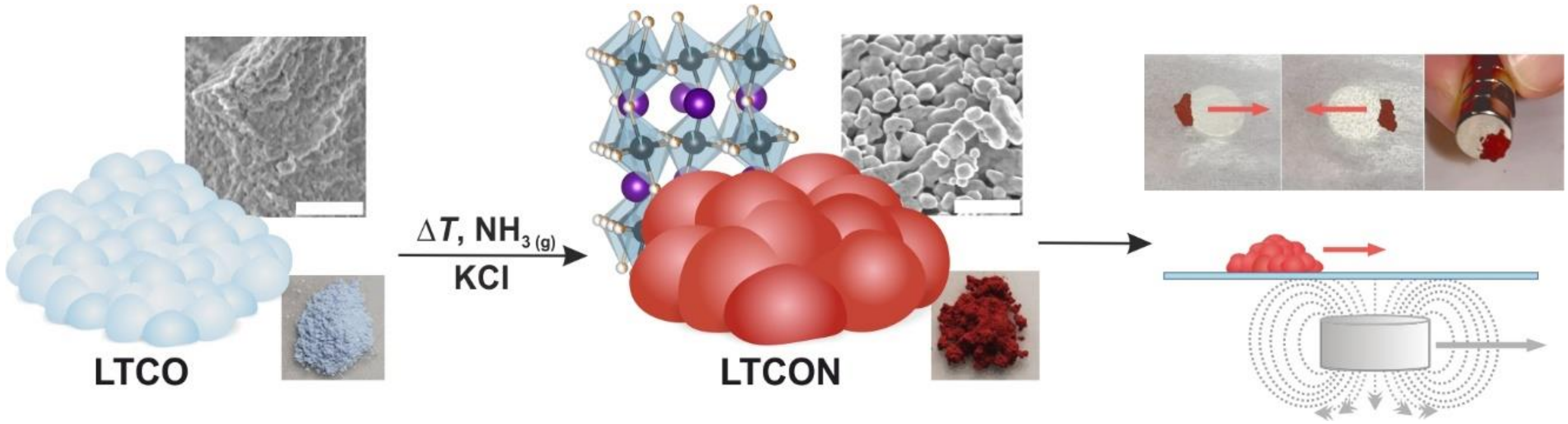

FIG. 1. Reaction path to LTCON. Reaction path from LTCO to LTCON, which can be attracted or moved by permanent magnets (Movie S1 and S2). Respective scanning electron microscopy (SEM) images (Scale bars: 500 nm) and sample photos are shown. In the depicted crystal structure, the purple ions represent $La^{3+}$. The Ta/Co ions (black/cyan) are 6-fold coordinated by $N^{3-}$ and $O^{2-}$ (white/orange) in an octahedral environment.

## III. STRUCTURAL AND COMPOSITIONAL INVESTIGATIONS OF LTCON

Scanning electron microscopy (SEM) shows that the synthesized LTCON particles exhibit an average size of 300 nm (Figs. 1, S1). However, high-resolution transmission electron microscopy (HR-TEM) reveal many smaller particles (average size 50 nm) which are agglomerated (Fig. 2a).

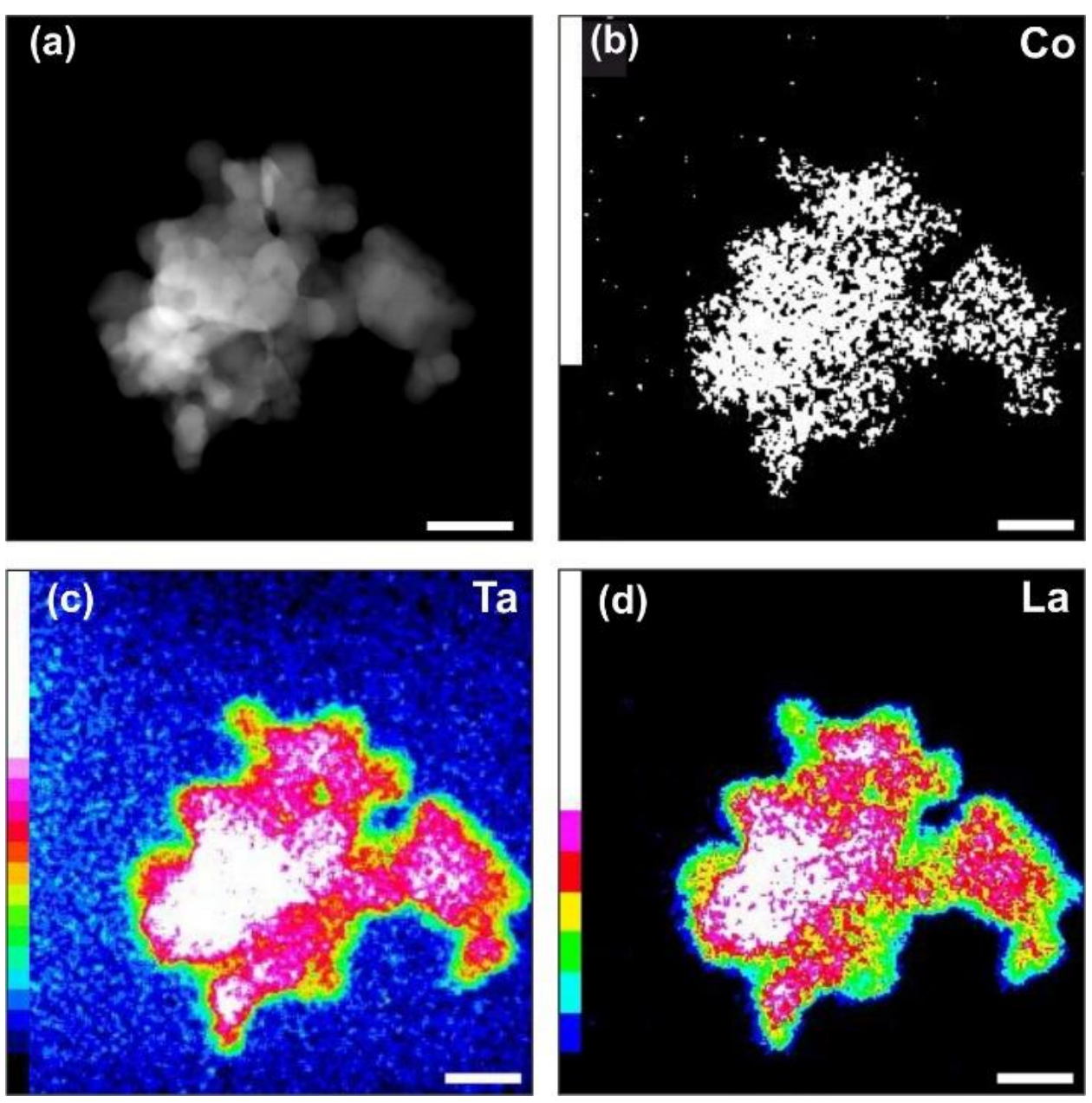

FIG. 2. HR-TEM investigations of LTCON-5. (a) Zoomed HAADF image of single-phase LTCON-5 nanoparticles and long-time measured EDX maps of (b) Co, (c) Ta, and (d) La. The different colors represent the concentration. By long-time measurements, the homogeneous distribution of the Co ions is obvious. Scale bars: 50 nm.

For imaging we used high-angle annular dark-field imaging (HAADF) and for further solubility investigations energy-dispersive X-ray spectroscopy (EDX). By conducting a long-time EDX measurement, the homogeneous distribution of all three elements (Ta, La, and Co) was verified. Especially, Co revealed a very homogeneous distribution in the material (Fig. 2b) without showing any sign of a secondary Co-rich phase in LTCON-5. For further information of the conducted electron microscopy experiments cf. Supplemental Material Sec. III.

The exact concentration of the cations (La, Ta, and Co ions) is investigated by inductively coupled plasma optical emission spectroscopy (ICP-OES) and reveals the expected compositions, which are implemented in Tab. S4. Furthermore, hot gas extraction (HGE) – determination of O and N contents – revealed anionic vacancies ($\delta$) for LTCON. We will further discuss the importance of anionic vacancies in Sec. VII.

Crystal structure analysis (Powder X-Ray diffraction combined with Rietveld refinements) revealed the space group type *Imma* (Tabs. S5–S22) for all LTCON samples. This is a perovskite-type phase where the Ta/Co ions are coordinated in an octahedral environment by six anions. The same space group type is already reported for the non-magnetic perovskite-type oxynitrides $LaTaO_2N$ and $LaTaON_2$ [17]. Therefore, a change in the space group type is not expected, because we used very small concentrations of Co ions for *B*-site substitution in $LaTa(O,N)_3$ which should not affect the crystal structure by itself.

The colors of the LTCON powders range from red for $x$ = 0.01 (LTCON-1) *via* dark red for $x$ = 0.03 (LTCON-3) to a very dark red for $x$ = 0.05 (LTCON-5) (Fig. 1 and Fig. S1). The obtained optical bandgap $E_G$ of the oxynitrides determined by diffuse reflectance spectroscopy (1.7 eV $\leq E_G \leq$ 1.9 eV) reflects the red color and points to a semiconductor (Fig. S8): metallic nanoparticles are typically black in color. Similar results in color and bandgap were obtained for pristine $LaTa(O,N)_3$ [17].

## IV. THE IMPORTANCE OF STOICHIOMETRY IN LTCON

As stated above, secondary phases containing magnetic Co, Fe or Ni ions can produce HT-ferromagnetism. Therefore, the main question in many studies is the solubility limit of such ferromagnetic ions in the materials matrix [2]. In principle, possible ferromagnetic secondary phases in our samples could be either Co-rich particles or elemental Co clusters/particles.

In order to investigate if elemental Co as secondary phase is possible in our case, we deliberately produced one reference sample of LTCON-5 containing a non-stoichiometric amount of Co ions ($Co^{z+}$ excess of 3.4 %)(cf. Supplemental Material Sec. III). In Figs. 3a–f, both samples – the stoichiometric and the Co-rich (non-stoichiometric) LTCON-5 – are presented and compared with each other. We used HAADF, EDX, and powder X-ray diffraction (PXRD) for thorough characterization of both samples (Figs. 3a-f). The contrast variation of the nanoparticles observed by the HAADF imaging in Figs. 3a,c,e can be attributed to different crystal orientations of the particles proven by electron diffraction (Fig. S14). EDX allows the investigation of the homogeneous distribution of elements in a material. First, we investigated the non-stoichiometric sample: in combination with EDX, HAADF and PXRD only Co(O,N) nanoparticles with 0.5 wt% as a secondary phase exhibiting a diameter of 40 nm $\leq d \leq$ 80 nm in the Co-rich LTCON-5 were found. Other elements such as La, Ta, O, and N, which were recorded, show a homogeneous distribution (Fig. S15 and Tab. S23). In Fig. 3f only PXRD reflections of the Co(O,N) phase in the Co-rich sample were found and no reflections of an elemental Co phase (Co-hcp or Co-fcc). This can be explained by the applied high temperatures and long ammonolysis periods [25] which do not allow the formation of elemental (metallic) phases. Therefore, elemental Co in the Co-rich LTCON-5 can be excluded.

Instead, by using the exact stoichiometric amount of Co ions for the synthesis neither Co(O,N) particles nor elemental

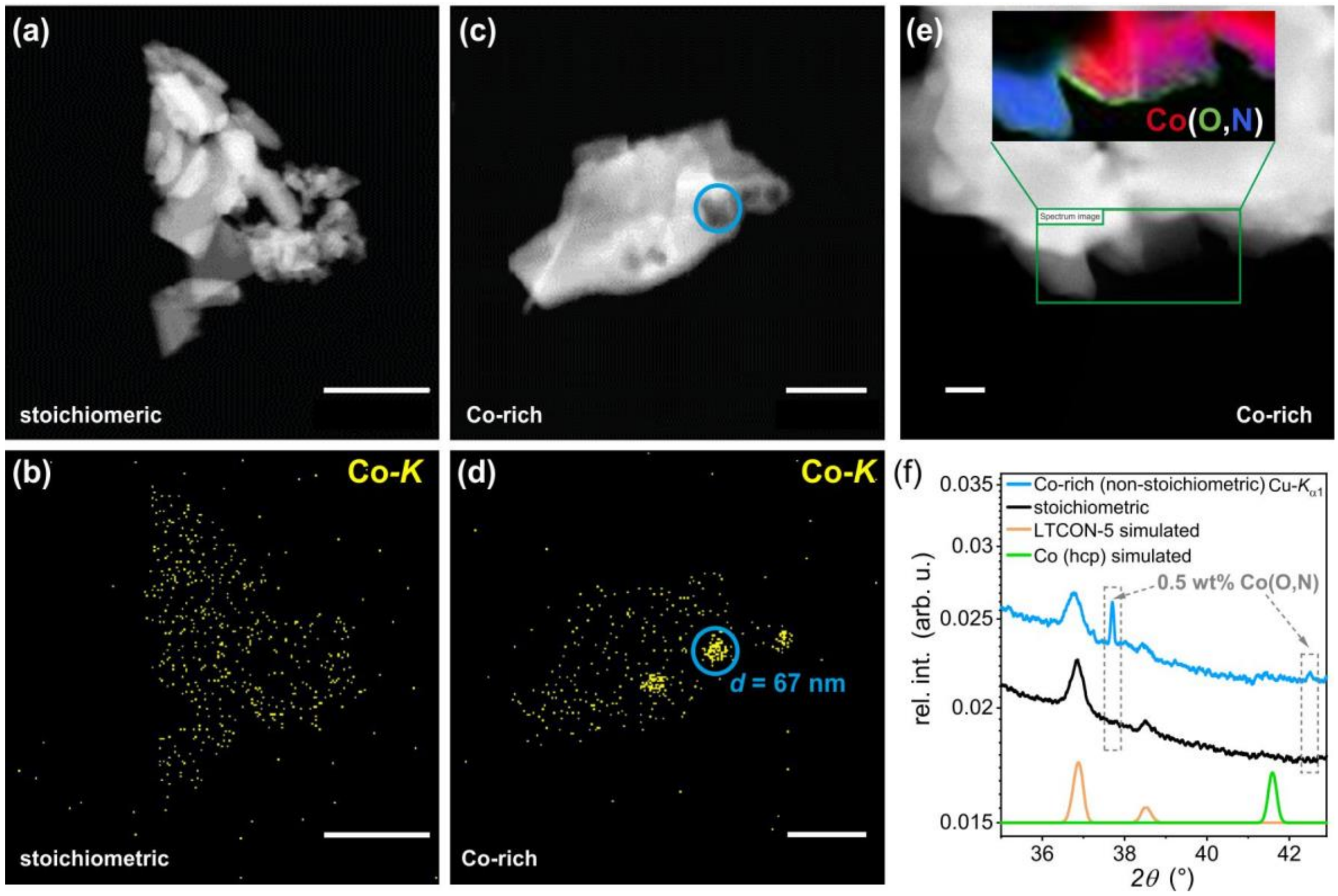

FIG. 3. HAADF/EDX and PXRD investigations of LTCON-5. (a)–(b) HAADF image (dark-field) and EDX map of single-phase (stoichiometric) LTCON-5 nanoparticles showing the homogeneous distribution of Co ions in the particles. Scale bars: 500 nm. (c) HAADF image of Co-rich LTCON-5 nanoparticles containing Co(O,N) nanoparticles. Scale bar: 250 nm. (d)–(e) EDX maps of Co-rich LTCON-5 nanoparticles containing Co(O,N) nanoparticles. Scale bars of (d): 250 nm and (e): 50 nm. In Fig. 3e, La and Ta are not measured. (f) PXRD data of the stoichiometric and Co-rich LTCON-5 samples.

Co particles as secondary phases are found (Figs. 3b and f). Hence, the possibility to obtain Co-containing secondary phases in a stoichiometric weighed sample by using the applied synthesis procedure is very unlikely. This demonstrates that we synthesized single-phase LTCON powders where the Co ions are completely included in the materials matrix.

## V. HT-FERROMAGNETISM IN SINGLE-PHASE PEROVSKITE-TYPE LTCON

Next, we investigated the magnetic properties of LTCON-1, LTCON-3, and LTCON-5 *via* superconducting quantum interference device (SQUID) measurements (Figs. 4a–c). At room temperature (300 K) a clear saturating behavior is observed, with 90 % of the saturation magnetization $M_s$ at fields of about 0.3 T. The insets in Figs. 4a–c reveal a clear hysteretic behavior (hysteresis loops) with coercive fields (Tab. S24) indicating ferromagnetism. At 300 K, $M_s$ increases with the amount of Co ions from LTCON-1 to LTCON-5 from 0.088 emu g$^{-1}$ to 1.179 emu g$^{-1}$ (Table I). In contrast to the observed clear HT-ferromagnetism of LTCON, the LTCO precursor exhibits just simple paramagnetism (Fig. S16) and pristine $LaTa(O,N)_3$ revealed at 300 K a diamagnetic behavior [17] in former studies. The processing of the raw SQUID data of LTCON is shown in Fig. S17. The observed strength of the ferromagnetic signal is orders of magnitude larger in contrast to former studies [6] (Supplemental Material Sec. IV), excluding measured SQUID artefacts and contaminants.

Additionally, all LTCON samples show an increase of a paramagnetic-like behavior with a similar absolute value at low temperatures. Because of the strong increase of the total sample magnetization with increasing Co ion concentration, the relative paramagnetic contribution is reduced for higher Co ion concentrations. This can be seen in the 2 K curves from LTCON-1 to LTCON-5 (Figs. 4a–c). The observed paramagnetism (Brillouin function shape) with respect to the ferromagnetism has a relative contribution of 60 % in LTCON-1. In comparison, for LTCON-5 the contribution difference is only 4 %. However, the absolute paramagnetic contribution is almost similar in strength for all three oxynitride samples.

TABLE I. SQUID-based magnetic parameters at 300 K. The extracted magnetic parameters are the saturation magnetization ($M_s$), and the magnetic moment ($m_{\text{Co ion}}$) per Co ion for each Co ion concentration ($x$).

| Compound | $x$ | $M_s$ (emu g$^{-1}$) | $m_{\text{Co ion}}$ ($\mu_B$/Co ion) |
|---|---|---|---|
| LTCON-1 | 0.01 | 0.088 | 0.60 |
| LTCON-3 | 0.03 | 0.543 | 1.28 |
| LTCON-5 | 0.05 | 1.179 | 1.53 |

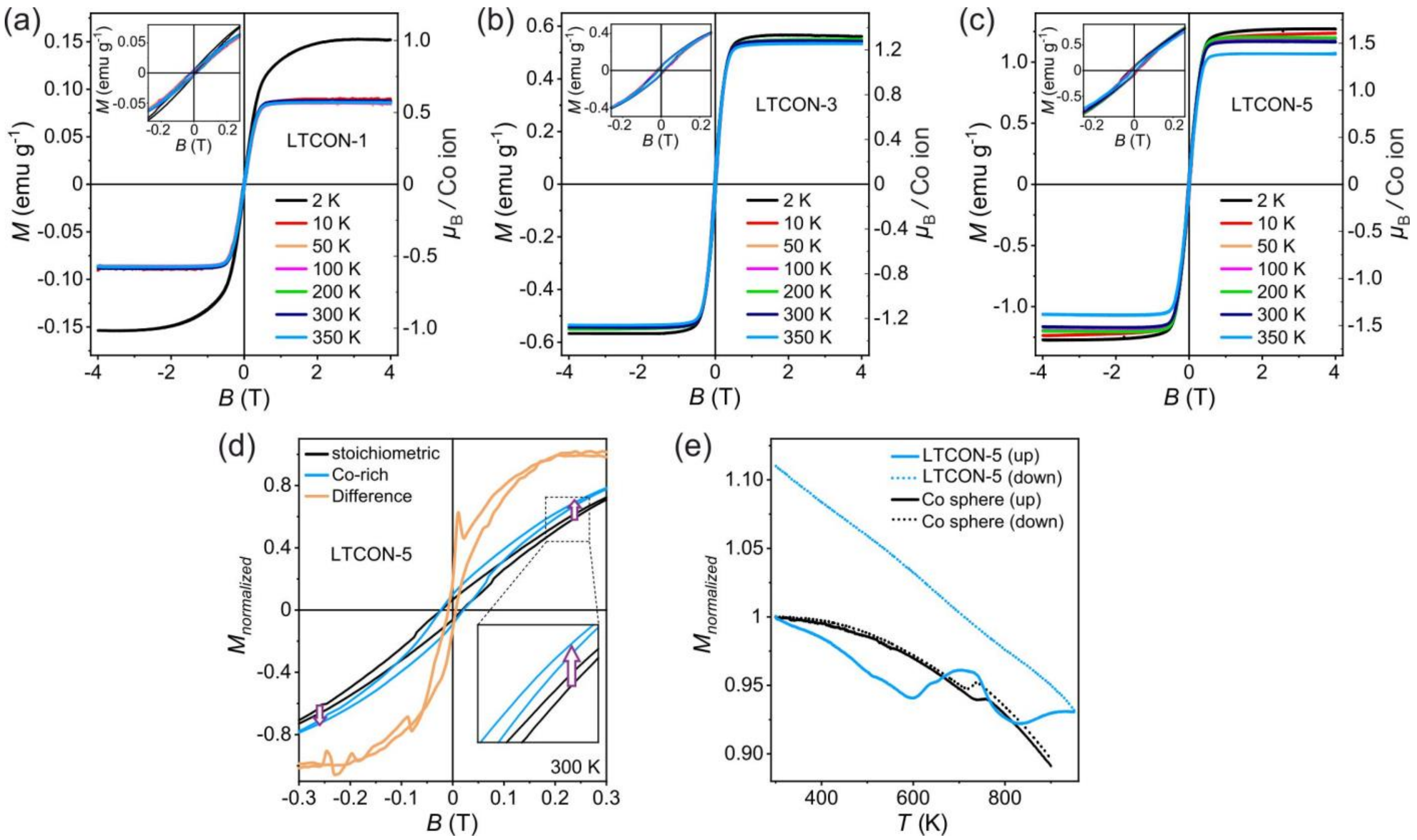

FIG. 4. Magnetic investigation of LTCON. (a)–(c) Magnetization curves of LTCON-1, LTCON-3, and LTCON-5. (d) Field-dependent magnetization curves at 300 K of the stoichiometric and Co-rich LTCON-5. The purple arrows indicate the almost constant difference between both. The difference has a much steeper slope at fields below 0.1 T. The difference signal could be related to the Co(O,N) particles identified in Fig. 3. For a better comparison, all curves were normalized. (e) Normalized magnetization versus temperature curves of single-phase LTCON-5 and a pure Co metal sphere.

In addition, a clear increase in the magnetic moment per Co ion is observed. If we attribute the whole sample magnetization completely to the given amount of Co ions, we can calculate a magnetic moment per Co ion. This is increasing from 0.6 $\mu_B$/Co ion to 1.53 $\mu_B$/Co ion. The observed unexpected high sample magnetizations lead to macroscopic attracting forces by conventional permanent magnets. Fig. 1 shows the mechanical movement of the whole LTCON-5 powder by permanent magnets (Movie S1 and S2). Of course, such magnetic force-related effects are always present for samples exhibiting the same magnetization. However – because the entire powder is moved – it demonstrates a homogeneous magnetic behavior combined with a homogeneous material composition.

In former reports undesired elemental transition metal clusters were responsible at least for some parts of the measured magnetization curves [8,26]. We already excluded particulate Co-rich secondary phases such as Co(O,N) and particulate elemental Co by chemical analysis. Now, we exclude elemental Co clusters by magnetic investigations because of their tiny size: small clusters of elemental Co would reveal a superparamagnetic behavior [27]. Therefore, we compared calculated magnetization curves of elemental Co clusters at different temperatures (10 K to 350 K) with the measured magnetization curve of LTCON-5 at 300 K (Fig. S18). The calculated magnetization curves for elemental Co clusters reveal a very strong temperature dependence indicating superparamagnetism (Supplemental Material Sec. V). Superparamagnetic materials also reveal a strong increase of the coercive fields at low temperatures, which remarkably decrease at room temperature [28,29]. Even larger elemental Co clusters [30] and elemental Co particles reveal observable temperature dependencies in saturation magnetization, shape, and coercive fields between 10 K and room temperature. All this cluster-like behavior is absolutely not observed for our samples (Figs. 4a–c): our obtained coercive fields and curve shapes are almost temperature-independent. Therefore, this by itself is a clear proof for the absence of superparamagnetic elemental Co clusters in LTCON.

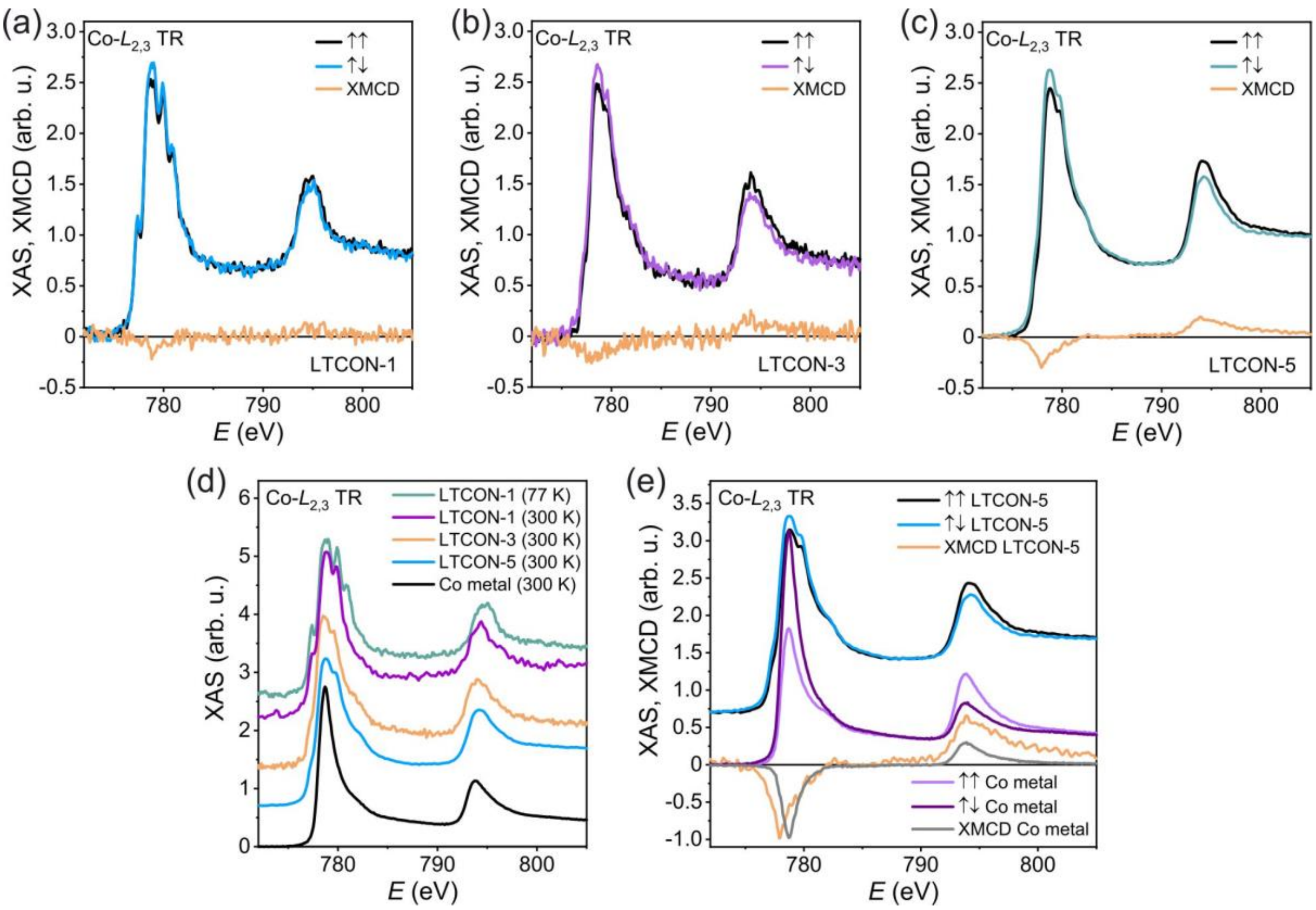

FIG. 5. XAS/XMCD measurements of LTCON. (a) TR spectrum of LTCON-1 measured at 77 K. (b) TR spectrum of LTCON-3 measured at 300 K. (c) Corresponding XAS/XMCD TR mode for LTCON-5 measured at 300 K. (d) Non-magnetic TR-mode XAS spectra of LTCON-1, LTCON-3, LTCON-5, and a pure metal Co TR spectrum measured at 300 K. In addition, a low temperature (77 K) TR spectrum of LTCON-1 containing more spectral details is presented. (e) TR-mode XMCD/XAS spectra of LTCON-5 and the Co metal reference sample. The XMCD signals were both scaled up to 1 in order to better compare the shape of both.

Not only the absence of temperature dependence, also the relative high saturation field is an additional proof for the absence of Co clusters. LTCON contains a very low Co ion concentration. Therefore, the average distance of potential tiny elemental Co clusters would be quite large. In comparison to Li *et al*. [29], the saturation field of Co clusters of 6 nm size with an elemental Co concentration of 43 ± 5 at% in the whole sample is at 0.15 T at room temperature. Since we have a Co ion concentration below 1 at%, we clearly can exclude dipole-dipole interactions between tiny Co clusters as a possible source for the wide hysteresis loops observed in LTCON.

Additionally, we compared the 300 K magnetization curves of the single-phase LTCON-5 and the Co-rich LTCON-5 (Fig. 4d). A clear difference is observed, that we attribute to the existing Co(O,N) nanoparticles in the Co-rich LTCON-5. We can estimate from the difference that the Co(O,N) particles have just a 7 % magnetic contribution to the total sample magnetization of the Co-rich sample. Furthermore, we performed SQUID measurements up to 950 K (Fig. 4e) in vacuum. The measurements show that the temperature dependence is quite different between LTCON-5 and elemental Co. The increase of the magnetization at 600 K indicates a partial decomposition by nitrogen loss and further vacancy formation rather than a complete decomposition of the material. It is worth to mention that this partial decomposition results in a 12 % increase in saturation magnetization, already indicating a correlation between ferromagnetism and vacancies. Since the $T_C$ seems to be higher than the observed partial decomposition temperature of the material and the crystal structure is thermodynamically stable over the measured temperature range in vacuum (Supplemental Material Sec. VI, Fig. S19), the $T_C$ cannot be determined with our SQUID system. In addition, the measured pure Co metal sphere shows the expected hysteretic hcp–fcc phase transition (local maximum), which is not visible in LTCON-5.

In order to identify possible different magnetic contributions of the Co ions to the sample magnetizations, we performed X-ray absorption spectroscopy (XAS) / XMCD at the Co-$L_{2,3}$ edges (Fig. 5). In Figs. 5a–c the bulk-sensitive transmission (TR) mode spectra of LTCON-1, LTCON-3, and LTCON-5 are presented. The measurements were performed at 300 K or at 77 K for LTCON-1. The 77 K spectrum of LTCON-1 is presented because of a more detailed peak structure originating from reduced Franck-Condon broadening [31]. By measuring LTCON-1 at 300 K (Fig 5d) the spectral shape resembles LTCON-3 and LTCON-5 at 300 K. In Figs. 5a-c, all three TR spectra reveal a significant XMCD signal and show similar multiplet-like peak structures giving the proof of oxidized Co species such as $Co^{3+}/Co^{2+}$ [32–34]. For other perovskite-type Co-containing oxides with $Co^{3+}$ or $Co^{2+}$ in octahedral coordination (*e.g.* $LaCoO_3$) also Co-$L_{2,3}$ edge spectra with very similar spectral shape were observed [32,35,36].

By comparing the TR spectra of LTCON with a Co metal reference sample (Fig. 5d), possible existing elemental (metallic) Co cluster and/or Co particles as the origin of the observed HT-ferromagnetism can be excluded, finally: The reference Co metal TR spectra are measured with the same measurement parameters (*e.g.* exactly the same energy resolution) at the same beamtime. This allows a direct and precise comparison of energy shifts and line shapes between all measured samples. All LTCON XAS spectra and especially LTCON-5 show a clear non-metallic broad multiplet-like peak structure with a distinct shoulder before and after the main peak. This is very different to the shape of the Co metal spectrum. More important is the fact that the obtained XMCD signal *e.g.* of LTCON-5 is much broader in shape and significantly shifted to lower photon energies compared to the Co metal XMCD spectrum (Fig. 5e). If we assume that the Co ions in LTCON-5 would be partially segregated to a secondary metallic phase like Co clusters or particles, and this phase would be only related to the observed ferromagnetism, one would expect almost exactly the same XMCD signal in position, shape, and width as the XMCD signal determined from the reference Co spectrum. This is not the case. Hence, the shape of the measured TR spectra of LTCON and the different XMCD signal to the XMCD signal of Co metal are directly proving the intrinsic ferromagnetism of LTCON. If we now assume that the solubility limit of Co would be below 5% at the Ta-site, the spectral shape should be quite different between LTCON-1 and LTCON-5, because the relative amount of Co metal-like spectra must increase for LTCON-5. However, all spectra look identically, proving that the solubility is higher.

By applying sum rules [37,38] on the TR spectrum of LTCON-5 with the “upper limit” (see below) of holes of $n_H = 4$ ($Co^{3+}$), a spin moment of $1.14 \pm 0.06$ $\mu_B/Co^{3+}$ and a vanishing orbital moment of $-0.01 \pm 0.01$ $\mu_B/Co^{3+}$ for LTCON-5 can be calculated. For LTCON-1 and LTCON-3, the determined moments are listed in Table II. Additionally, by applying sum rules on the reference Co metal spectrum, a spin moment of $m_s = 1.64 \pm 0.08$ $\mu_B/Co^0$ and an orbital moment of $m_l = 0.16 \pm 0.008$ $\mu_B/Co^0$ are determined. These values are perfectly in agreement with the moments observed in well-known literature such as Chen *et al.* [39] and prove the accuracy of the determined sum rule results for LTCON in Table II. The XMCD-based magnetic moments for all LTCON powders are 25 % below the SQUID-based magnetic moments per Co ion. The measured TR spectra point to IS-$Co^{3+}$ consistent to the obtained 25% reduced magnetic moments in comparison to the SQUID results [36]. However, other oxidation state contributions such as $Co^{2+}$ cannot be excluded. Therefore, $n_H$ (*e.g.* by using $n_H = 3$ for $Co^{2+}$) will be below 4 on average for all Co ions and, hence, the sum rule results further decrease. Because of the observed difference to the SQUID-based values, the obtained XMCD results proof that besides the ferromagnetic Co ions in LTCON the host matrix has to be magnetically polarized. This can also explain the ferromagnetic long-range ordering which is observed in LTCON.

TABLE II. Orbital moments and spin moments of LTCON. The moments were determined *via* the measured XMCD/XAS spectra.

| Compound | $x$ | $m_{orb.}$($\mu_B/Co^{3+}$) | $m_{spin}$ ($\mu_B/Co^{3+}$) |
|---|---|---|---|
| LTCON-1 | 0.01 | $-0.001 \pm 0.01$ | $0.44 \pm 0.02$ |
| LTCON-3 | 0.03 | $-0.002 \pm 0.01$ | $1.0 \pm 0.05$ |
| LTCON-5 | 0.05 | $-0.01 \pm 0.01$ | $1.14 \pm 0.06$ |

If we still would try to attribute the sample magnetizations to a non-observed secondary phase, by questioning the observed spectral shapes and the energy shifts between Co metal and LTCON XMCD spectra, the total magnetization must then be related to this secondary phase. As we have demonstrated above for the non-stoichiometric sample we can easily observe a secondary phase with a $Co^{z+}$ excess of 3.4 %. Thus, we can start a thought experiment: We imagine a two-phase sample with 10 % from a ferromagnetic Co-containing phase and 90 % from a non-ferromagnetic LTCON phase (Tab. S27). In this case the Co magnetic moment would belong completely to the ferromagnetic secondary phase. If we now take the result of the Co magnetic moment for the whole LTCON-5 sample ($1.14 \pm 0.06$ $\mu_B/Co^{3+}$) we have to multiply by 10 the magnetic moment of $Co^{3+}$ to get the magnetic moment for just 10 % of a ferromagnetic Co-containing secondary phase because of linear correlation. This means a Co spin moment of 11.4 $\mu_B/Co^{3+}$ would be present in the 10 % ferromagnetic secondary phase, whereas 0 $\mu_B/Co^{3+}$ would be in the 90 % of non-magnetic LTCON-5. The result of 11.4 $\mu_B/Co^{3+}$ is not possible.

To conclude, by having a close look at all results, even if a very small Co metal-like phase would be present it cannot explain the existing XAS/XMCD results and electron microscopy studies. All results tend to an intrinsic ferromagnetism of LTCON originating from diluted Co ions in the material.

## VI. COMPARISON TO A SIMILAR SYSTEM

For comparison we synthesized a similar system, namely $LaTa_{0.95}Ni_{0.05}(O,N)_{3-\delta}$ (LTNON). The Ni-based material exhibited the same space group type (*Imma*) and red color as LTCON (Supplemental Material Sec. VII, Fig. S20 and Tabs. S25, S26). However, by applying crystal structure analysis we found a very low solubility limit for Ni ions in the material (≤ 0.6 at%) revealing secondary phases such as $Ni_3N$, also easily observable as Ni-rich regions in the EDX measurement (Fig. S21). Additionally, the not-single-phase LTNON exhibited a small ferromagnetic contribution in combination with a large temperature-dependent paramagnetic contribution (Fig. S22) underlining the results that $Ni^{z+}$ was not completely included into the crystal matrix (secondary phase visible). Since the chemical behavior of $Ni^{z+}$ differs from that of $Co^{z+}$ we want to emphasize that every other ion exhibits chemically different behaviors to $Co^{z+}$ leading to different compositions and solubility limits. Therefore, our conclusion is that $Co^{z+}$ is currently the only ion to get our observed results.

Hence, we demonstrated that it requires a careful evaluation of the synthesis parameters such as cationic stoichiometry and the choice of the right magnetic ion ($Co^{z+}$) for the synthesis to obtain a ferromagnetic behavior above RT in a DMS material.

## VII. DISCUSSION OF THE HT-FERROMAGNETISM IN LTCON

As the presence of HT-ferromagnetism in LTCON is highly unexpected we try to provide a preliminary discussion of this phenomenon in comparison to the previous explanations given for the room temperature ferromagnetism found in pristine ZnO. For this compound, surface-related vacancies were identified to explain ferromagnetic coupling and the magnetic volume fraction was shown to be related to the amount of non-stoichiometric grain boundaries [12,14,15]. This is also consistent to recent studies on highly defective 2D ZnO nanosheets [40]. Since LTCON contains a significant amount of anion vacancies (Tab. S4), one could suggest that the effect observed in LTCON even in the bulk could be similar to the vacancy dominated ferromagnetic foam as discussed in other $d^0$ ferromagnetic materials [12,14,15]. Additionally, the introduced nitrogen ion can play a significant role by varying the ionic and covalent character of the *B*–*X* bonding [22], which might support the vacancy introduced changes in the electronic structure. Considering this, the mechanism of the observed HT-ferromagnetism in LTCON seems to be more complicated than that of defect-ridden ZnO.

The magnetization increases by a factor of about 13, whereas the Co ion concentration is increased by a factor of 5 (Table I and II): this is directly related to the XMCD-based change in $Co^{3+}$ magnetic moments. This suggests that the Co ion is also mediating ferromagnetism and demonstrates a clear key ingredient. Since the calculated magnetic moments per Co ion determined from the XMCD measurements are at least 25 % lower than the by SQUID-determined values, we conclude that all of the existing Co ions are not sufficient to explain the observed magnetic behavior. This again supports the presence of a bulk-like ferromagnetism, whereas besides the Co ions extra magnetic moments must exist.

One also might think about a thread-like 3D single Co ion percolation network. If this network would be present, due to statistical Ta-site occupation a significant fraction of Co ions should not contribute to the percolation network, and a larger fraction of paramagnetic ions should exist in the bulk, which is also not observed in LTCON.

Hence, after two decades of intensive research in the magnetism community, our study demonstrates the realization of a HT-ferromagnetic DMS by partially substituting Ta with Co ions in a $LaTa(O,N)_3$ matrix. This highly reproducible result lays the foundation for a new research field investigating and developing same or similar DMS systems. Furthermore, it can resurrect the general research field of DMS, which had fallen out of favor over the last few years due to the lack of success. We expect that future investigations will focus on discovering new material matrices, which exhibit room temperature DMS, and understanding the underlying origin of the dilute ferromagnetic ordering.

## ACKNOWLEDGMENTS

The authors thank Mr. Samir Hammoud for HGE and ICP-OES measurements, Dr. Sebastian Bette, Mrs. Christine Stefani, Prof. Dr. Dinnebier for PXRD (D8 Bruker) measurements, Mrs. Annette Fuchs and Prof. Dr. Joachim Maier for nitrogen sorption, Dr. Maximilian Hackner and Prof. Dr. Joachim Spatz for providing the glovebox for synthesis. For fruitful discussions and proof reading we acknowledge Dr. Max T. Birch. Thanks goes to Dipl.-Ing. Claudia Fasel and Dr. Rotraut Merkle for TGA-MS measurements and Dr. Sven Fecher and Dr. Songhak Yoon for fruitful discussions. The authors acknowledge the financial support of the Institut Laue Langevin, Grenoble, France and the reactor beamtime. We thank the synchrotron light source KARA and the KNMF, both Karlsruhe, Germany, for the provision of beamtime. C.B., M.W. and A.W. thank the German Research Foundation for financial support within the priority program SPP 1613 "Solar $H_2$" (WE 2803/7-1).

## APPENDIX

### 1. Methods

#### *a. Synthesis of LTCO*

The oxides were prepared *via* a Pechini method which is based on the one already reported for n-LTO [17]. First, $TaCl_5$ (Alfa Aesar, 99.99 %) and $Co(NO_3)_2 \cdot 6H_2O$ (Merck, EMSURE®) were loaded in an exact stoichiometric amount into a Schlenk flask under argon and 50 mL of dried methanol was added. The amount of substance of the *B*-site cations Ta and Co was calculated to be 0.01 mol in total. For the Co-rich sample an additional amount (4 %) of $Co(NO_3)_2 \cdot 6H_2O$ was added. Afterwards, 0.03 mol water-free citric acid (Sigma Aldrich, ≥ 99.0 %) for all samples was added. 0.01 mol $La(NO_3)_3 \cdot 6H_2O$ (Sigma Aldrich, 99.99 %) was weighed into a second Schlenk flask and citric acid was added in the same molar ratio as for the *B*-site cations. The mixture was dissolved in 10 mL of dried methanol and the solutions of both Schlenk flasks were combined in one Schlenk flask and stirred under reflux for 2 h at 353 K with addition of a 15-fold molar excess of ethylene glycol (Merck, EMPLURA®). The dispersion was transferred to a crystallizing dish and heated for 10 h at 393 K followed by a thermal treatment at 573 K for 5 h as reported before [17]. The resulting black xerogel was calcined in an alumina crucible for 16 h at 923 K to obtain the nanocrystalline $LaTa_{1-x}Co_xO_{4-\delta}$ (LTCO) with $x$ = 0.01 (LTCO-1), $x$ = 0.03 (LTCO-3), $x$ = 0.05 (LTCO-5).

#### b. Synthesis of LTCON

The synthesis of $LaTa_{1-x}Co_x(O,N)_{3-\delta}$ (LTCON) with $x$ = 0.01 (LTCON-1), $x$ = 0.03 (LTCON-3), $x$ = 0.05 (LTCON-5) is described in the following: 350 mg of LTCO was loaded into an alumina boat and placed into a conventional thermal gas flow ammonolysis setup. Ammonolysis was carried out for 10 h at 1,223 K with two subsequent ammonolysis cycles for 14 h at 1,273 K and KCl flux addition (1:1 weight ratio). The applied ammonia flow for all cycles was 300 $mL{\cdot}min^{-1}$ $NH_3$ (Westfalen AG, > 99.98 %). The same conditions were used to prepare the Co-rich LTCON-5 sample.

### 2. Sample Characterization

#### *a. Crystal Structure*

Powder X-ray diffraction (PXRD) measurements at room temperature were carried out on a Bruker D8-Advance powder X-ray diffractometer using Cu-$K_{\alpha_1}$ radiation (Ge(111) monochromator), Bragg Brentano geometry, and a Lynx-Eye detector. Additionally, a Rigaku Smartlab powder X-ray diffractometer (Cu-$K\alpha_{1,2}$) was used. The continuous scans covered an angular range of $5° \leq 2\theta \leq 90°$ with an angular step interval of 0.007°. The collected diffraction data were evaluated *via* Rietveld refinements using *FullProf.* 2k [41–43]. Pseudo-Voigt functions were selected to describe the reflection profile and the background was linearly interpolated between a set of background points with refinable heights.

#### *b. Chemical Composition*

The chemical composition of the produced samples was investigated *via* inductively coupled plasma optical emission spectroscopy (ICP-OES) using a Spectro Ciros CCD ICP-OES instrument for cations and hot gas extraction technique (HGE) using an Eltra ONH-2000 analyzer for the anions.

#### *c. In Situ Experiments - Formation*

*In situ* ammonolysis [17] was performed by thermogravimetric analysis (TGA) using a Netzsch STA 449 F3 Jupiter. The measurements were carried out under flowing $NH_3$ (80 $mL{\cdot}min^{-1}$ $NH_3$ + 8 $mL{\cdot}min^{-1}$ Ar) with a heating rate of 10 $K{\cdot}min^{-1}$ up to 1,273 K.

#### *d. X-Ray Photoelectron Spectroscopy*

X-ray photoelectron spectroscopy (XPS) was carried out using a Kratos Axis Ultra system with a monochromatized Al-$K_a$ source (1,486.6 eV) holding a base pressure in the lower $10^{-10}$ mbar range. The powders were fixed on an indium foil and a flood gun was used in order to avoid charging effects. The binding energy was calibrated by setting the C 1*s* of adventitious carbon to 284.5 eV [44] with respect to the Fermi level. Analysis of the XPS data was performed with Casa XPS software. The energy separation and peak area of the Ta $4f_{7/2}$ and Ta $4f_{5/2}$ orbitals were constrained according to literature [44]. The low signal to noise ratio together with a non-flat background did not allow for any reasonable fitting of the Co 2*p* region.

#### *e. Electron Microscopy Investigations*

The particle morphology of the produced LTCON and LTCO was analyzed *via* scanning electron microscopy (SEM) (ZEISS GeminiSEM 500, 2 kV) and the in-lens detector was used for imaging. For energy-dispersive X-ray spectroscopy (EDX) a window-containing XFlash® 6|60 (Bruker) detector was used. The accelerating voltage for EDX measurements was 15 kV.

For transmission electron microscopy (TEM) investigations the particles of LTCON were dispersed in ethanol and drop-cast on a Cu grid covered with an amorphous carbon foil and with a mesh size of 200 μm provided by Plano. Small accumulations of oxynitride and oxide particles, respectively, were investigated on a Philips CM-200 FEG TEM operated at 200 kV, applying bright- and dark-field imaging. In order to verify the space group determined by Rietveld refinements of the PXRD and neutron diffraction (ND) data, selected-area diffraction patterns were recorded. The recorded polycrystalline diffraction patterns were analyzed by using the JEMS software package [45]. Colored grain orientation maps were constructed by the overlay of five dark-field images recorded with varying beam tilt. The composition and homogeneity of selected particles were analyzed with an EDX system from EDAX. Elemental mappings were collected with a probe size of 3 nm, a step size of ~2 nm, and a dwell time of 15 ms per pixel. The objective polepiece of the device contains only iron.

High-resolution transmission electron microscopy (HR-TEM) was performed by using the state-of-the-art JEOL ARM200F TEM coupled with an EDX system from JEOL (plus Bruker) at the "Stuttgart Center for Electron Microscopy (StEM)" in Stuttgart. The microscope was operated at 200 kV. The LTCON-5 particles were prepared as described above for the Philips CM-200 FEG TEM. To discriminate the measured cobalt concentration from the cobalt in the polepiece, Co-free $LaTaON_2$ (synthesized after ref. [17]) was measured as reference.

#### *f. Diffuse Reflectance Spectroscopy*

UV-visible diffuse reflectance spectra (DRS) were recorded by using a Carry 5000 UV–VIS NIR spectrophotometer. The spectra were measured in the range of 200 nm to 800 nm and the Kubelka-Munk [46] conversion was applied to the recorded reflectance spectra. The optical bandgaps were estimated by extrapolating the onset of absorption to the abscissa.

### *g. SQUID*

Magnetometer surveys were carried out with a commercial VSM MPMS3 superconducting quantum interference device (SQUID) from Quantum Design. This system allows both conventional DC and VSM-type measurements. The temperature ranged from 1.8 K up to 350 K (oven option: $T <$ 1,000 K, $p <$ 150 mTorr), while the field was switched up to 4 T. For zero field cooling purposes the magnet was quenched to minimize the residual magnetic field. Depending on the sample and the measurement type the effective sensitivity is in the range of $10^{-8}$ – $10^{-9}$ emu. The pressed powder anisotropy measurements for in- and out-of-plane measurements were performed with a Quantum Design MPMS 7 system, because the detection system is less sensitive for variations in the filling factor, providing better precision and comparability for anisotropic sample geometries. The weight of the measured pure Co metal sphere was 4.3 mg. The diamagnetic background originating from the sample holder was subtracted from a finite linear slope determined from the negative slope of the raw SQUID data (Fig. S17). Since perovskite-type oxynitride powders exhibit a very bad sinter ability (therefore high grain-boundary resistivity) and the particles are in the nm-range, carrier-mediated magnetic measurements were not performed.

### *h. XAS / XMCD*

X-ray magnetic circular dichroism (XMCD) and X-ray absorption spectroscopy (XAS) experiments were performed at the synchrotron ANKA/KARA at KIT, Karlsruhe, in order to measure local atomic magnetic moments. All XMCD and XAS spectra were recorded at the WERA beamline with an energy resolution of about $\Delta E/E = 2\cdot 10^{-4}$. At the Co–$L_{2,3}$ edge the degree of circular polarization was 82 %, which was used for sum rule corrections. We have used our own superconducting magnet end station providing ultra-fast field switching with ramping rates up to 1.5 $T\cdot s^{-1}$. All spectra were measured in an applied magnetic field up to 4 T. Each XMCD spectrum was measured as a function of energy with fixed light helicity and field. The energy was swiped uniformly with a rate (monochromator speed) of 0.2 $eV\cdot s^{-1}$ measured for each spectrum while simultaneously reading out the transmission data (TR) and incident X-ray (Au grid for $I_0$) current with about one data point every 0.03 eV. In order to get the XMCD data the circular light helicity was chosen (R = right, L = left), then two measurements were performed with reversed magnetic field (N = north; S = south). The sequence for a single full step run was RN→RS→LN→LS→LN→LS→RN→RS. This sequence minimized the effects of drift and further possible systematic errors. For better statistics, final spectra were averaged over consecutive spectra (in both helicities). TR and $I_0$ were measured using Keithley 6517A electrometers. The ramping rate (0.2 $eV\cdot s^{-1}$) was carefully chosen that no observable energy broadening could be detected. Each single XAS spectrum took about 3-5 minutes. No noticeable energy drift was observed between consecutive single spectra. Non-magnetic XAS spectra were obtained by averaging the magnetic XAS spectra for parallel and antiparallel aligned light helicity vs. magnetization direction. For the XMCD measurements, the sample powders (35 mg) were dispersed in 150 µL of ethanol (≥99.5 %, Ph. Eur.) and 50 µL of a terpineol (Aldrich, water-free) / ethyl cellulose (#46070, Fluka) / ethyl cellulose (#46080, Fluka) / ethanol (≥99.5 %, Ph. Eur.) mixture. The mixtures (“glue”) were prepared as follows: first, a 10 wt% solution in ethanol (≥99.5 %, Ph. Eur.) of both ethyl cellulose batches was prepared. Afterwards, both 10 wt% ethyl cellulose solutions (#46070, and #46080), terpineol, and ethanol were mixed together in a 2.2 : 2.8 : 2.1 : 1.5 weight ratio. The dispersion of particles, ethanol, and “glue” was drop-casted on a 5 × 5 mm “SiN” membrane (thickness of 100 nm) which was sputtered with a 5 nm Cr thin film for better adhesion and electrical conductivity. As a detector for the transmitted light an almost magnetic field-independent Hamamatsu GaAsP diode (G1116 type) was used.

### *i. Neutron Diffraction*

Neutron diffraction was carried out on the high-resolution D2B diffractometer ($\lambda$ = 1.59417(2) Å) of the Institut Laue Langevin (ILL). The diffractograms of LTCON were recorded at 10 K and 300 K. The ND data (DOI: 10.5291/ILL-DATA.EASY-471, 10.5291/ILL-DATA.EASY-472, 10.5291/ILL-DATA.6-06-482) was refined using the *FullProf* 2k [43] and a pseudo-Voigt function was chosen to generate the line shape of the reflections. No magnetic refinements of the data were possible as due to the low amount of Co present in the samples (Ta was assumed to be non-magnetic) any magnetic contribution to the scattering falls short of the detection limit of neutrons.

### *j. Nitrogen Sorption*

In order to investigate the specific surface area of the oxide precursors nitrogen sorption was carried out using an Autosorb-1-MP (Detection limit: $S_{BET} >$ 1 $m^2\cdot g^{-1}$) from Quantachrome Instruments. First, the samples were annealed at 393 K in order to remove adsorbed water. Adsorption and desorption isotherms were collected at 77 K. To determine the specific surface area the Brunauer-Emmett-Teller [47] (BET) method was used.

### *k. TGA-MS*

To determine possible adsorbed water and organic residues on the oxide precursors surface TGA coupled with mass spectrometry (MS) was carried out on a Netzsch STA 449 C Jupiter coupled with a QMS 403C Aeolos® mass spectrometer.

The oxide was heated to 1,473 K at a rate of 5 K·min$^{-1}$ under synthetic air and then cooled to room temperature.

# Supplemental Material:

## Observation of a possible diluted ferromagnetism above room temperature in cobalt-substituted $LaTa(O,N)_{3-\delta}$

Cora Bubeck[†], Eberhard Goering[†], Robert Lawitzki, Kathrin Küster, Wilfried Sigle, Marc Widenmeyer, Ulrich Starke, Clemens Ritter, Gabriel J. Cuello, Peter Nagel, Michael Merz, Stefan Schuppler, Gisela Schütz & Anke Weidenkaff

† These authors contributed equally to this work and should be considered as co-first authors.

## Supplemental Material – Discussion

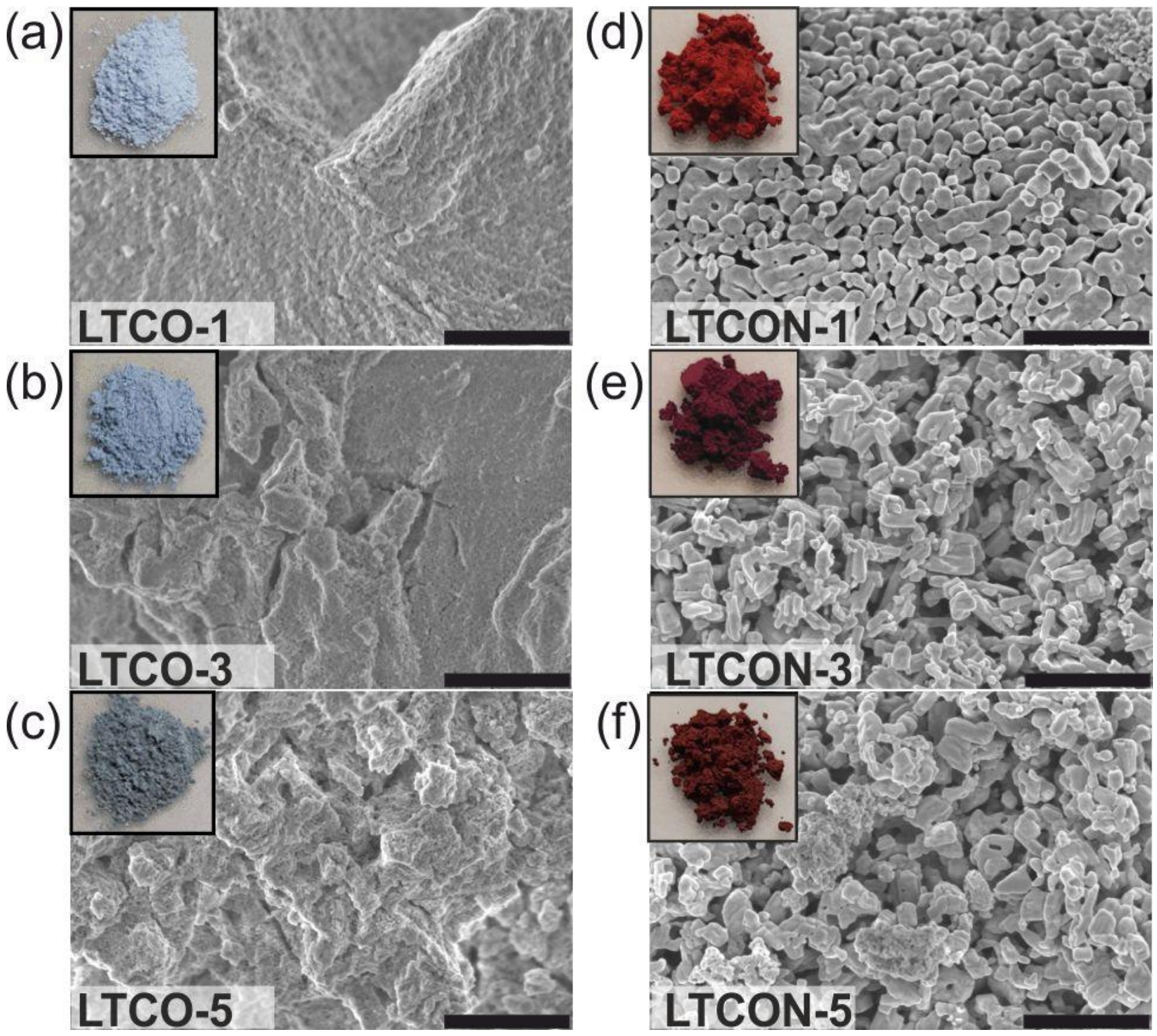

**Fig. S1.** SEM images, and sample photos of the oxides and oxynitrides. Respective SEM images of (a)–(c) all LTCO and (d)–(f) all LTCON powders showing primary particles in the nm-range with the respective sample photos showing the colored powders. Scale bars: 1 µm.

## Supplemental Material Sec. I: Chemical analysis of LTCO

Powder X-ray diffraction (PXRD) reveals nanocrystalline oxide precursors for all three Co ion concentrations (Fig. S2). Similar PXRD patterns were already obtained for the synthesized and used nanocrystalline oxide precursors in ref. [1]. Because of the nanocrystallinity of the oxides, the determination of phase purity was not possible by PXRD. Therefore, the oxides were investigated by transmission electron microscopy (TEM) and energy-dispersive X-ray spectroscopy (EDX) (Fig. S3). The obtained TEM images show tiny primary particles in the nm-range, which were agglomerated to porous secondary particles pointing to the nanocrystallinity of the powders. By recording selected area electron diffraction (SAED) patterns no rings were visible. The recorded EDX maps reveal for all three Co ion concentrations a homogeneous distribution of Co and Ta and give no hint for secondary phases. The resolution was 10 nm per single Co photon count. Hence, we assume because of the EDX and SAED results the oxides to be phase pure.

To determine the anionic and cationic compositions of the single-phase oxides hot-gas extraction (HGE) (Tab. S1) and inductively coupled plasma optical emission spectroscopy (ICP-OES) were used, respectively. The lower obtained oxygen weight fractions for the oxides in comparison to the expected ones with a ratio of La:(Ta,Co):O = 1:1:4 ($x$ = 0.01: 16.73 wt% O, $x$ = 0.03: 16.83 wt% O, $x$ = 0.05: 17.31 wt% O) were pointing to oxygen vacancies in the samples (Tab. S1). The expected oxygen weight fractions for the oxides were derived from the very similar assumed ratio for n-LTO (nanocrystalline "$LaTaO_4$") which was obtained at the same temperature with the same synthesis method. [1] By calcinating LTCO-1 ($LaTa_{0.99}Co_{0.01}O_{3.99(6)}$) at 1273 K to form crystalline LTCO-1 HGE measurements revealed a composition of $LaTa_{0.99}Co_{0.01}O_{3.93(9)}$. The found amount of oxygen vacancies in crystalline LTCO-1 (c-$LaTa_{0.99}Co_{0.01}O_{3.93(9)}$) seems plausible from the fact that Co as *B*-site cation has a lower valence state than Ta which has normally 5+ [1] resulting in a higher concentration of anionic vacancies because of charge compensation. In order to verify the phase purity of crystalline LTCO-1 a PXRD pattern was recorded (Fig. S4). The recorded pattern was refined in the space group type $Cmc2_1$ revealing a single-phase oxide because no additional reflections were observed (Tabs. S2–S3).

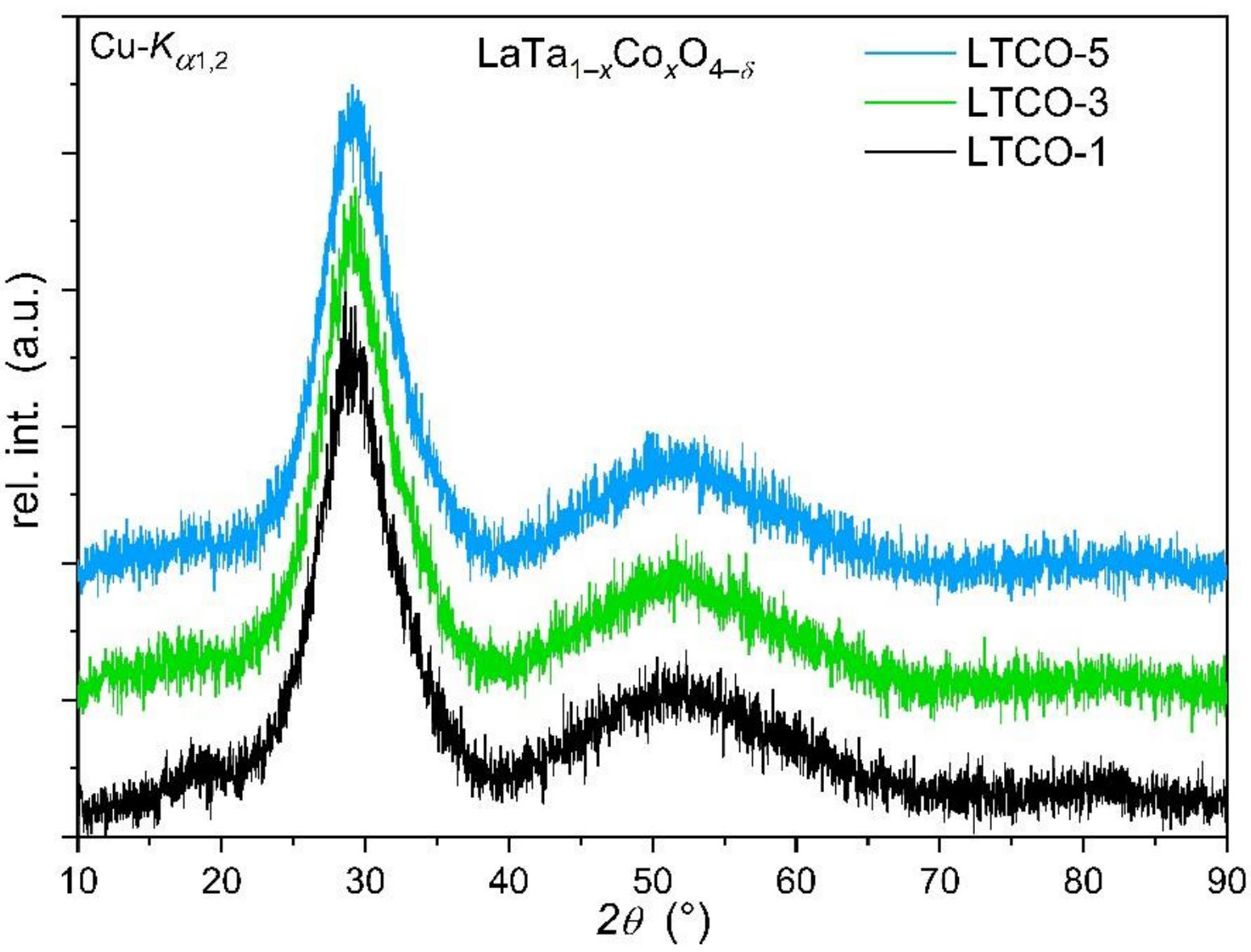

**Fig. S2.** Powder X-ray diffraction (PXRD) patterns of nanocrystalline LTCO. The patterns show nanocrystalline, nearly amorphous $LaTa_{1-x}Co_xO_{4-\delta}$ (LTCO) obtained after the used soft chemistry procedure. The anionic compositions were determined by HGE (Tab. S1).

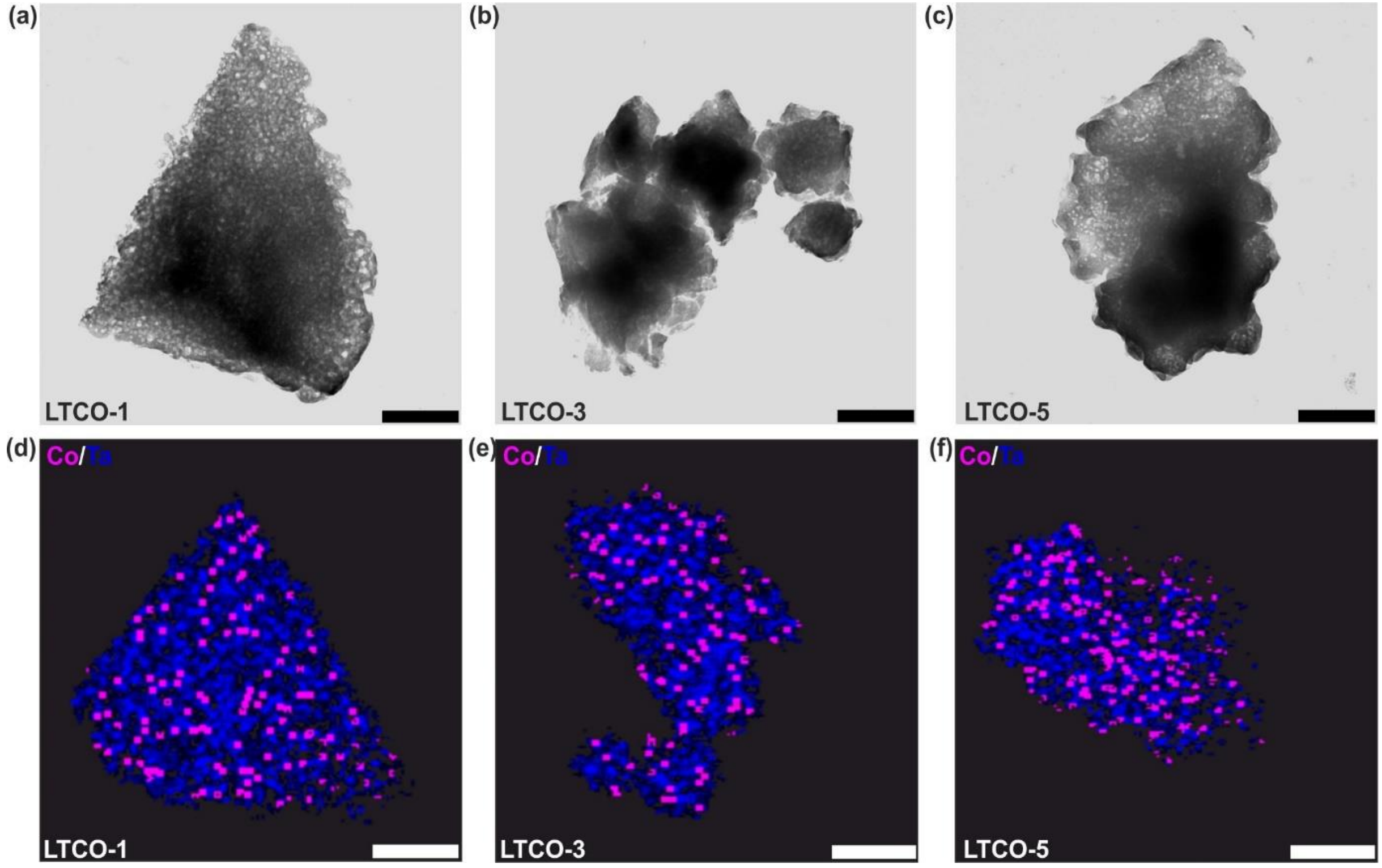

**Fig. S3.** TEM images and EDX mappings of LTCO. (a)–(c) TEM images of particles of the oxides and (d)–(f) EDX maps of the LTCO powders showing the homogeneous distribution of Co (pink) and Ta (blue) in the particles. The pixels were averaged with 3 x 3 pixels showing a pixel size of $d$ = 10 nm. This means that one Co ion is present in a region of 10 x 10 nm² because of the low number of single counts. Scale bars of all images: 100 nm.

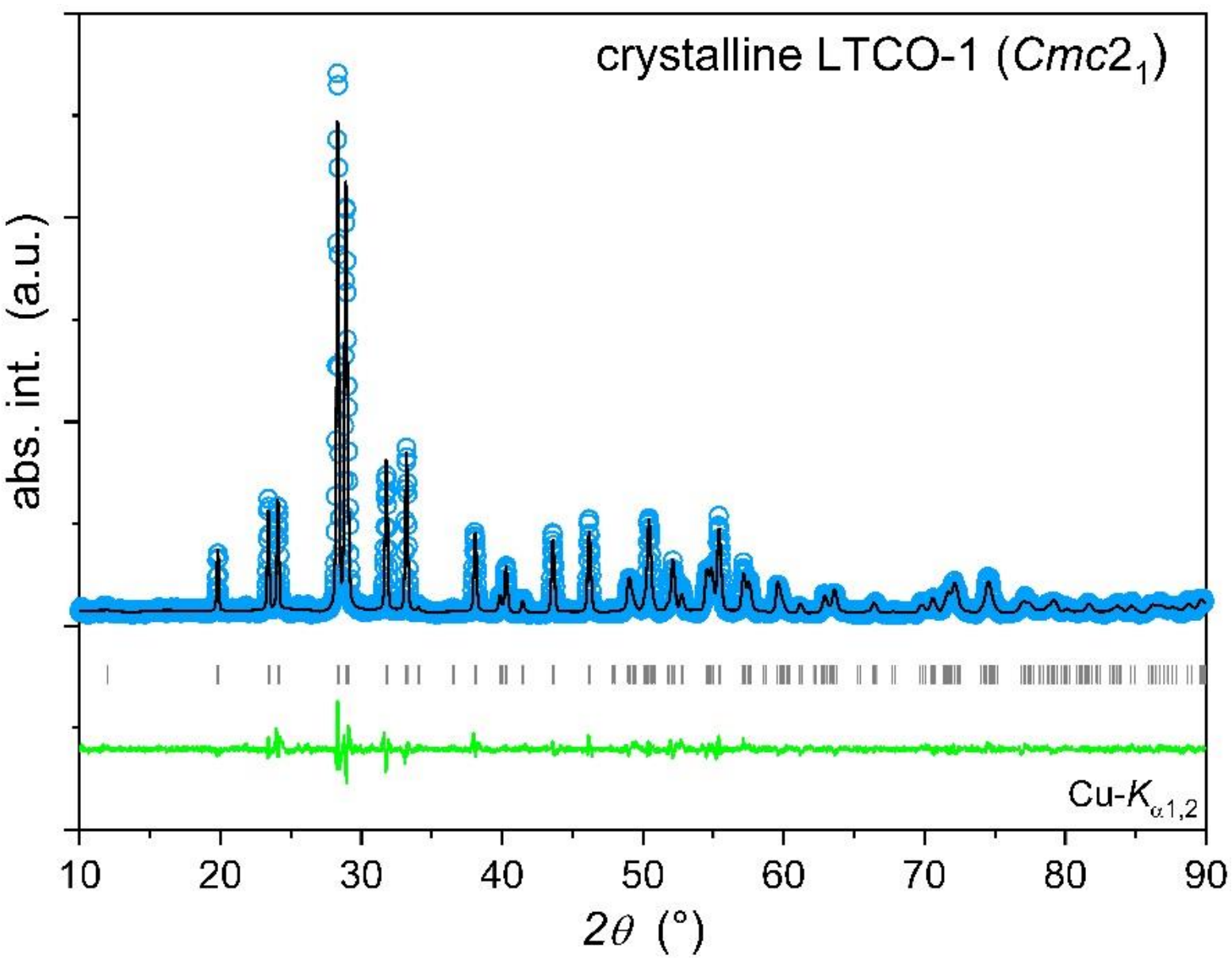

**Fig. S4.** Rietveld refinements of crystalline LTCO-1.The refinements reveal space group type *Cmc*$2_1$ for crystalline LTCO-1 (c-$LaTa_{0.99}Co_{0.01}O_{3.93(9)}$).

**Tab. S1.** Compositions of the oxide precursors. Oxygen content of LTCO and crystalline LTCO-1 (calcined at 1273 K) determined by hot gas extraction (HGE).

| **Compound** | $w(O_{expected})$ (wt%) | $w$(O) (wt%) | O |
|---|---|---|---|
| **LTCO-1** | 16.73 | 16.7 ± 0.3 | 3.99(6) |
| **LTCO-3** | 16.83 | 16.7 ± 0.3 | 3.97(1) |
| **LTCO-5** | 16.94 | 16.6 ± 0.4 | 3.91(9) |
| **crystalline LTCO-1** | 16.73 | 16.5 ± 0.3 | 3.93(9) |

**Tab. S2.** Unit cell parameters of crystalline LTCO-1 (c-$LaTa_{0.99}Co_{0.01}O_{3.93(9)}$).They were determined by Rietveld refinements.

| **Unit Cell Parameter** | **crystalline LTCO-1** |
|---|---|
| **Space group type** | $Cmc2_1$ |
| $a$ **(Å)** | 3.9304(4) |
| $b$ **(Å)** | 14.7690(1) |
| $c$ **(Å)** | 5.6286(5) |
| $V_{cell}$ **(Å³)** | 326.73(5) |
| $\rho$ **(g·cm⁻³)** | 7.785 |
| **Phase fraction (wt%)** | 100 |
| $R_p$ **(%)** | 7.45 |
| $R_{wp}$ **(%)** | 9.50 |
| $\chi^2$ | 2.26 |
| $R_{Bragg}$ **(%)** | 4.51 |

**Tab. S3.** Refined atom positions of crystalline LTCO-1 (c-$LaTa_{0.99}Co_{0.01}O_{3.93(9)}$).The determined space group type is $Cmc2_1$.

| **Atom** | **Wyck. Symb.** | $x$ | $y$ | $z$ | $B_{iso}$ **(Å²)** | **sof**[1] |
|---|---|---|---|---|---|---|
| La | 4*a* | 0 | 0.1699(2) | 0.01956[2] | 1.6(8) | 1[2] |
| Ta | 4*a* | 0 | 0.4146(2) | 0.05882[2] | 2.1(4) | 0.99[2] |
| Co | 4*a* | 0 | 0.4146(2) | 0.05882[2] | 2.1(4) | 0.01[2] |
| O(1) | 4*a* | 0 | 0.281(2) | 0.25982[2] | 0.5[2] | 1[2] |
| O(2) | 4*a* | 0 | 0.350(2) | 0.81404[2] | 0.5[2] | 1[2] |
| O(3) | 4*a* | 0 | 0.513(3) | 0.40898[2] | 0.5[2] | 1[2] |
| O(4) | 4*a* | 0 | 0.888(2) | 0.13644[2] | 0.5[2] | 1[2] |

[1]site occupancy factor, [2]fixed

**Supplemental Material Sec. II: Chemical analysis and formation of LTCON**

The cationic and anionic compositions of the oxynitrides were determined by HGE and ICP-OES (Tab. S4). The results revealed anionic vacancies for all three Co ion concentrations and, hence, were leading to the following anionic compositions: $LaTa_{0.99}Co_{0.01}O_{0.96(3)}N_{1.98(2)}\square_{0.05(5)}$ (LTCON-1), $LaTa_{0.97}Co_{0.03}O_{0.99(1)}N_{1.97(3)}\square_{0.03(4)}$ (LTCON-3), $LaTa_{0.95}Co_{0.05}O_{1.12(6)}N_{1.84(1)}\square_{0.03(3)}$ (LTCON-5). The square (□) represents the vacancies.

The crystal structures of the obtained oxynitrides were proven by neutron diffraction (ND) and powder X-ray diffraction (PXRD). Rietveld refinements of the PXRD and ND data reveal the space group type *Imma* for all oxynitrides (Figs. S5–S6, and Supplementary Tables 4–22) which was already observed for the perovskite-type oxynitrides $LaTaON_2$ and $LaTaO_2N$. [1] The contributions of the vanadium can (V) and cryo furnace (CF) (at 10 K) were considered in the refinements of the ND data at 10 K and 300 K since, small sample amounts of ca. 600 mg were used for measurements. Due to the small Co ion concentrations in the samples' magnetic contributions in the ND data was neither found at 10 K nor at 300 K.

To investigate the oxidation states of Ta in LTCON X-ray photoelectron spectroscopy (XPS) measurements were carried out and revealed two binding characters: $Ta^{III}$–(O,N) and $Ta^{V}$–(O,N) (Fig. S7). The XPS investigation and determination of the binding characters of the oxynitrides was done according to ref. [1]. The value of the binding energy $E_{B,\mathrm{Ta}4f7/2} = 23.7$ eV which refers to $Ta^{III}$–(O,N) according to the applied point charge model [1,2] was also found during *in situ* ammonolysis of n-LTO [1]. XPS is a very surface-sensitive method because of the very small mean free paths of the electrons [3]. Therefore, the concentrations of the Ta oxidation states reflects the first few nanometers and, hence, not the whole sample. [1] Consequently, $Ta^{3+}$ is just present at the surface whereas $Ta^{5+}$ is the main oxidation state. This was already observed for $LaTa^{V}ON_2$ [1]. The chemical environment of Ta in LTCON was supposed to be similar to $LaTa(O,N)_3$ [1]. This can be better understood by thinking about the potential Co ion concentration per unit cell in LTCON. For example, in one unit cell of LTCON-5 (formula unit $FU = 4$) just 0.2 Co ions on average are present leading to a similar chemical environment of Ta ions in LTCON-5 as in $LaTa(O,N)_3$. The determined Ta oxidation states indicated to oxidized Co including several possible oxidation states such as $Co^{2+}$ and $Co^{3+}$. In order to get an additional information about the oxidation state of the Co ions in the samples we performed XPS measurements. However, the amount of Co ions ($\leq 1$ at%) is too low for detection. XPS is a very surface-sensitive method, and hence, the amount of possible detected Co ions is even lower. Therefore, additional information of the Co oxidation states in the samples cannot be obtained. Therefore, the only indication of the Co oxidation states can be obtained by X-ray absorption spectroscopy (XAS), where *e.g.* $Co^{2+}$ and $Co^{3+}$ with similar spectral shape as reported in ref. [4] can be found.

For LTCON-1 the red color is very similar to the previously reported color of $LaTaO_2N$ [1] indicating a low optically active defect concentration [1]. However, diffuse reflectance spectroscopy (DRS) measurements reveal an increasing reflectance ($R$) beyond the optical bandgap to higher wavelengths by increasing Co ion concentration (Fig. S8a). The optical bandgaps of LTCON were determined by applying the tangent method on the to Kubelka-Munk [5] plots converted curves as for $LaTa(O,N)_3$ [1] (Fig. S8b). The values of the optical bandgaps are: 1.9 eV (LTCON-1), 1.8 eV (LTCON-3), and 1.7 eV (LTCON-5). The measured DRS revealed the characteristic shape and behavior of a typical pure semiconductor absorption edge, for example in ZnO [6,7] or $TiO_2$ [8], while the spectra of metallic precipitates are quite different, even for very small metal bulk contributions [9,10].

The formation of LTCON was exemplarily investigated by *in situ* ammonolysis of LTCO-1 (Fig. S9a). From 298 K to 738 K a first mass change of –0.5 % can be derived from the desorption of organic residues and adsorbed water from the sample which was reported before for *in situ* ammonolysis of nanocrystalline lanthanum tantalum oxide (n-LTO) [1]. To confirm this, a thermogravimetric analysis coupled with mass spectrometry (TGA-MS) (Fig. S10) in synthetic air was carried out and revealed besides the measured mass change (–0.5 %) weak MS signals for $m/z$ = 17 ($OH^+$), 18 ($H_2O^+$), 44 ($CO_2^+$). At $T$ > 1120 K a second mass change of –3 % was observed caused by a further release of $CO_2$ ($m/z$ = 12 ($C^+$), 22 ($CO_2^{2+}$), 44 ($CO_2^+$), 45 ($^{13}CO_2^+$)) and some $NO_2$ ($m/z$ = 46 ($NO_2^+$)) remaining from the used nitrate starting materials. This confirms the before stated assumption. Since during *in situ* ammonolysis ammonia as reducing atmosphere was present, a further formation of oxygen vacancies in LTCO-1 can take place besides the release of $H_2O$, $CO_2$, and nitrous gases. The formation of oxygen vacancies during ammonolysis of oxide precursors was already observed for $LaTaO_4$ [1]. According to ref. [1] the following negative mass change, which occurs at 878 K (–1.6 %) in the sample is the starting point of nitrogen incorporation. Our LTCO-1 precursor already contains oxygen vacancies like the termination product of microcrystalline $LaTaO_4$ during *in situ* ammonolysis before nitrogen incorporation [1]. Additionally, it has a similar microstructure as the reported n-LTO [1] – tiny particle size of several nm, nanocrystalline, high specific surface area ($S_{BET}$) of 31.3 m²/g. Therefore, we assume a product between $LaTaON_2$ [1] (obtained from $LaTaO_4$) and $LaTaO_2N$ [1] (obtained from n-LTO). This can also be verified by the "intermediate" reaction progression of LTCON in Fig. S11 (inset). After 10 h at 1223 K an overall negative mass change of –5.0 % was achieved. This is smaller than calculated for a full conversion to the expected product $LaTa_{0.99}Co_{0.01}ON_2$ (expected mass change: $\Delta m_{calc.}$ = –5.2 %) meaning the conversion was not completed after the first cycle. During cooling as for n-LTO [1] an absorption and desorption of gaseous species was observed. In order to identify the received product after the first ammonolysis cycles the formation of LTCON-1 and LTCON-3 were investigated by *ex situ* ammonolysis. After the first *ex situ* ammonolysis cycle at 1223 K / 10 h both products of LTCON-1 and LTCON-3 in comparison to the products of n-LTO and m-$LaTaO_4$ ($LaTaO_2N$ and $LaTaON_2$) [1] and in agreement with the observed progression of the *in situ* ammonolysis of LTCO-1 are not phase pure (Figs. S9b,c). Instead, a secondary phase in both compounds (shoulders at the reflections for LTCON-1 and a clear crystalline secondary phase for LTCON-3) was observed. The secondary phase for LTCON-3 shows a similar pattern as the crystalline LTCO-1 in Fig. S4, which can be one factor for the intermediate reaction progression. Additionally, the continuing mass decrease observed during the first (–5 %) and second cycle (–1.4 %) of *in situ* ammonolysis along with the total mass change of –6.4 % confirms the formation of a perovskite-type phase, keeping in mind the observed release of organic residues. The obtained reaction behavior of Co-substituted samples clearly deviated from the reported behavior of n-LTO [1]. On the one hand, a crystalline oxide precursor ($LaTaO_4$) with particles in the µm-range and low specific surface area ($S_{BET}$) leads to the O : N ratio of 1 : 2. On the other hand, a nanocrystalline oxide precursor (n-LTO) with particles in the nm-range and higher $S_{BET}$ leads to O : N = 2 : 1 [1]. In our case, we obtained "$LaTa_{1-x}Co_xON_2$" as a product in-between containing particles with a size in the nm-range instead of µm-range but the absence of $Ta^{4+}$. We assume, that the oxygen vacancies present in our oxide LTCO precursors are the driving force for this. Also, the present Co ions can support the enhanced vacancy formation. In that case, the already present anionic vacancies can force an enhanced vacancy formation and nitrogen incorporation leading to a higher nitrogen content and with it to a ratio of O : N = 1 : 2. Additionally, the present oxygen vacancies can hamper the Ta reduction since by X-ray photoelectron spectroscopy (XPS) measurements a $Ta^{III}$–(O,N) and a $Ta^{V}$–(O,N) binding character (instead of a $Ta^{IV}$–(O,N) binding character) was determined for all three oxynitrides (Fig. S7). This can be understood by the higher nitrogen content forcing the Ta to stay in the 5+ state for charge compensation. Therefore, we can add besides the recent reported microstructure of the oxide precursor [1] the presence of oxygen vacancies in the oxide precursor as parameter to tailor the oxidation state of *B*-site cations in oxynitride phases.

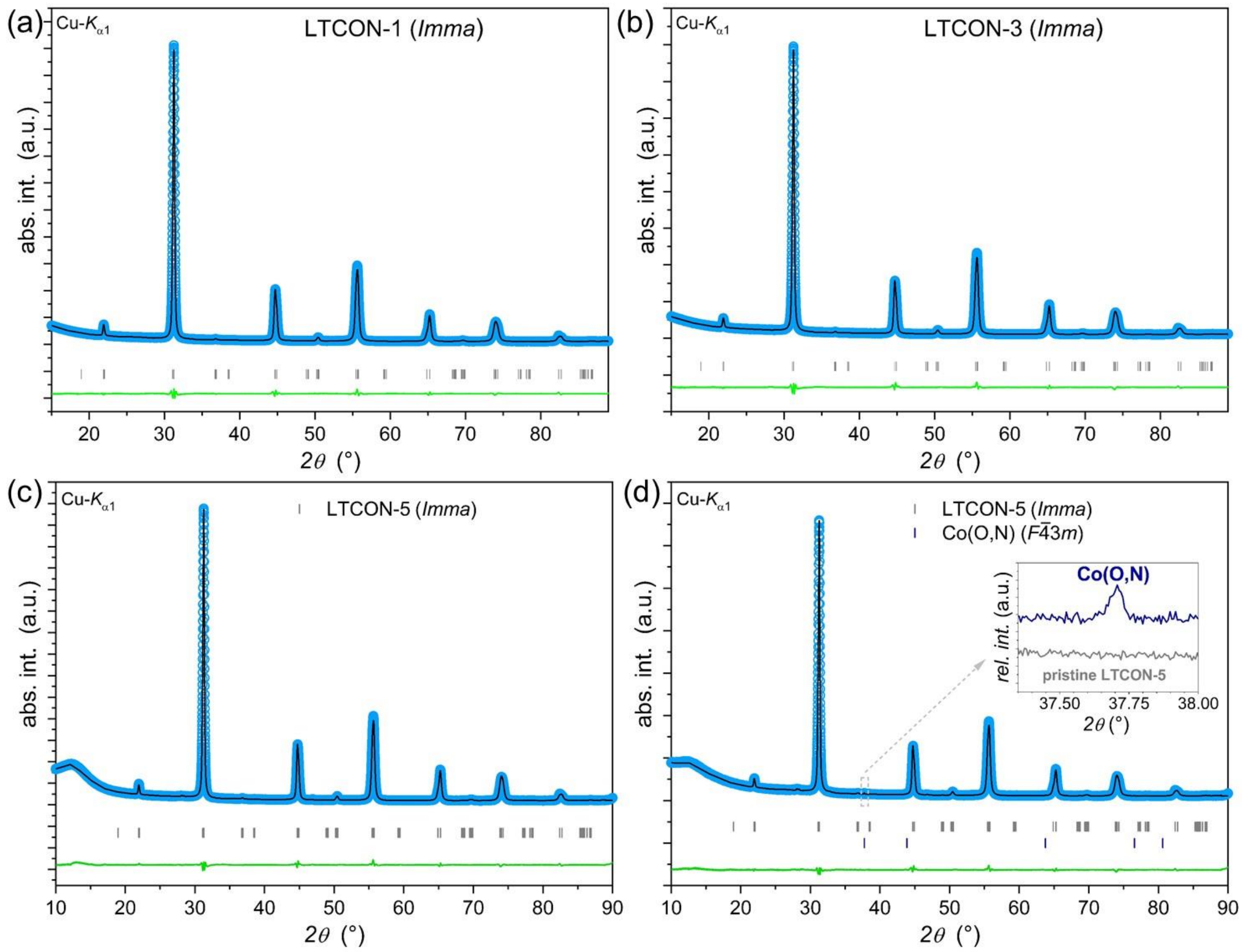

**Fig. S5.** Rietveld refinements of the PXRD data of the oxynitrides. (a)–(c) Rietveld refinements of PXRD data of single-phase LTCON-1, LTCON-3, and LTCON-5 at 300 K in space group type *Imma*. (d) Rietveld refinements of LTCON-5 containing Co(O,N) as secondary phase.

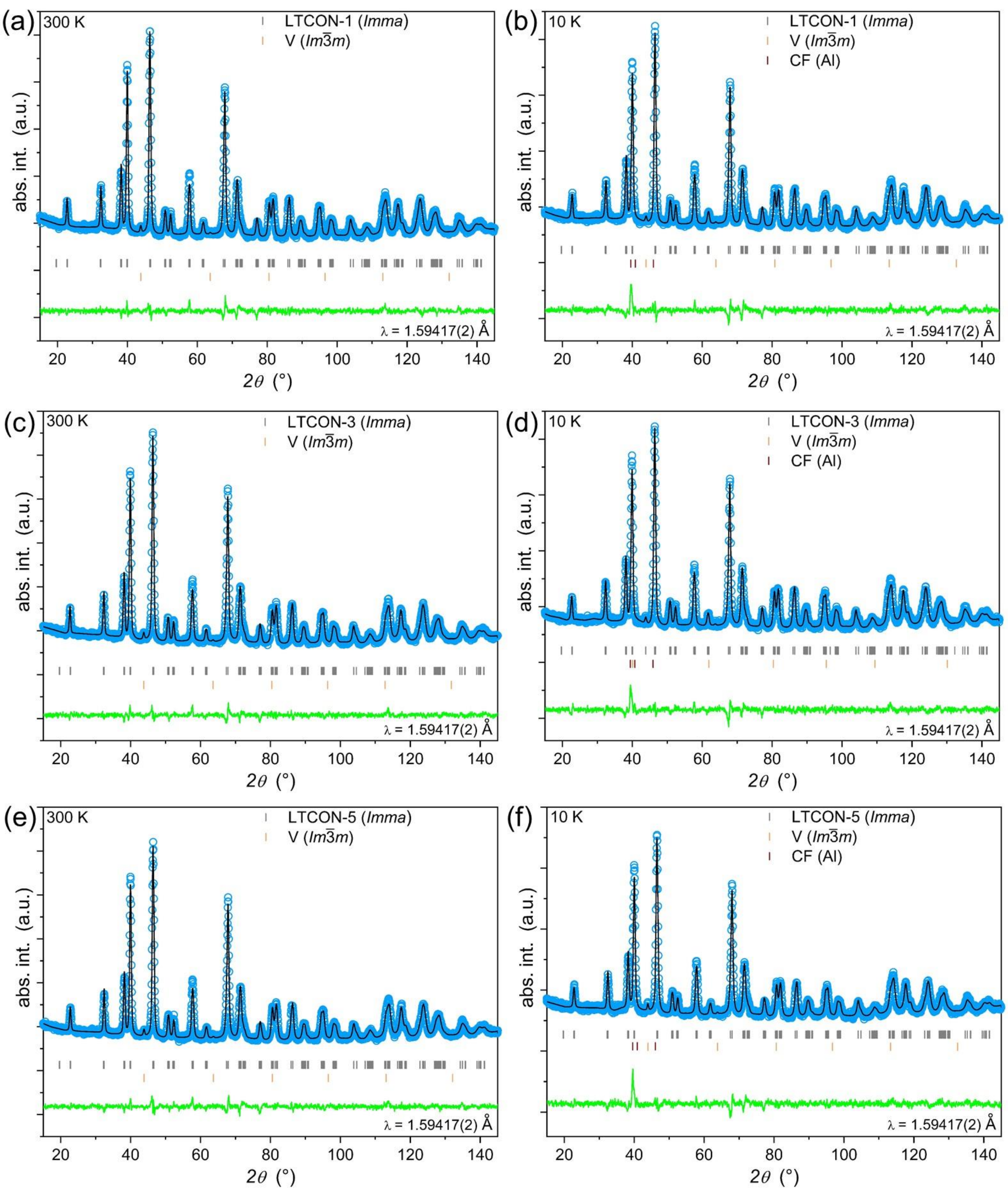

**Fig. S6.** Rietveld refinements of the ND data of LTCON. Rietveld refinements of ND data of (a) LTCON-1 at 300 K in space group type *Imma*. (b) LTCON-1 at 10 K in space group type *Imma*. (c) LTCON-3 at 300 K in space group type *Imma*. (d) LTCON-3 at 10 K in space group type *Imma*. (e) LTCON-5 at 300 K in space group type *Imma*. (f) LTCON-5 at 10 K in space group type *Imma*. CF refers to the scattering contribution of the cryo furnace.

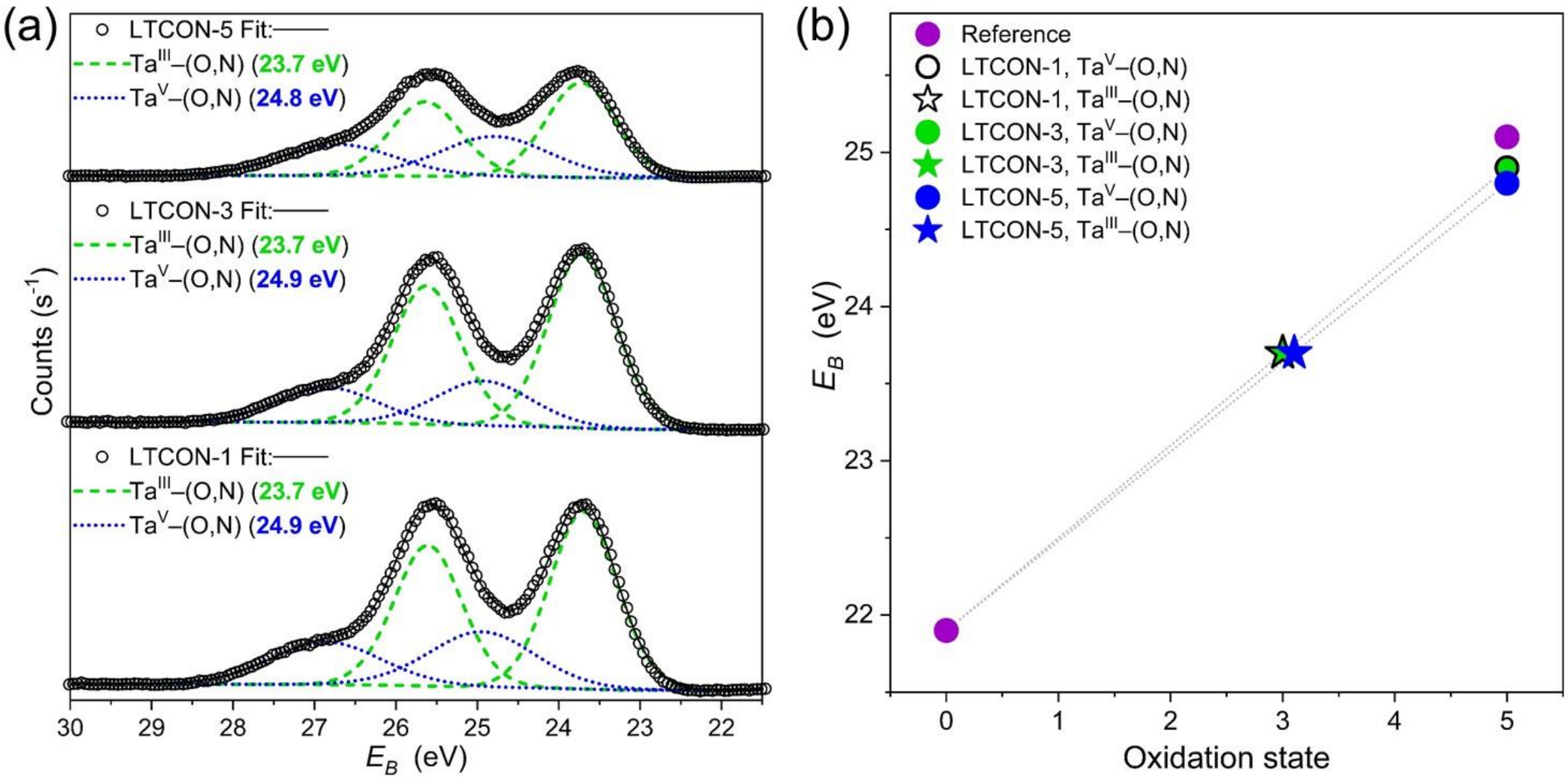

**Fig. S7.** XPS measurements of the Ta 4*f* regions of the oxynitrides. (a) Ta 4*f* spectra of LTCON and (b) the applied point charge model [1,2]. The binding energy of the $Ta^{5+}$ reference as reported in ref. [1] was used.

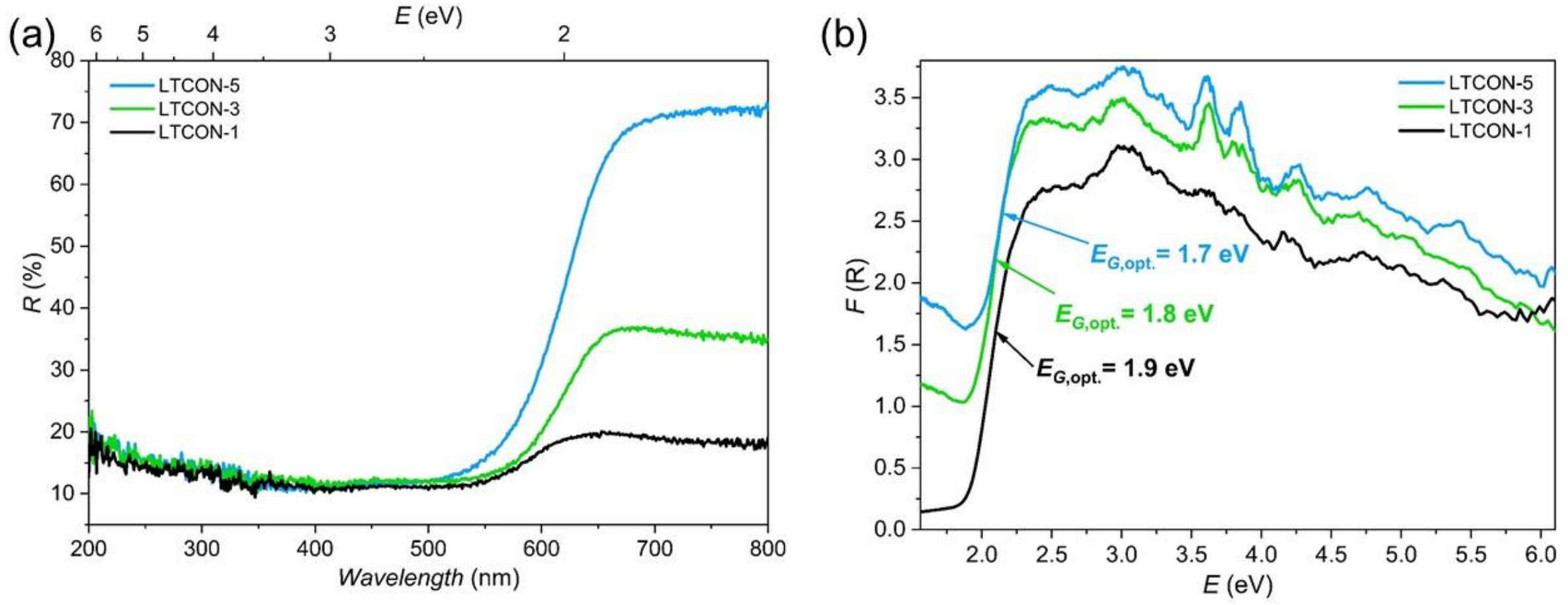

**Fig. S8.** DRS measurements of LTCON. (a) Measured reflectance spectra of LTCON and (b) Kubelka-Munk converted curves of the measured reflectance spectra for LTCON.

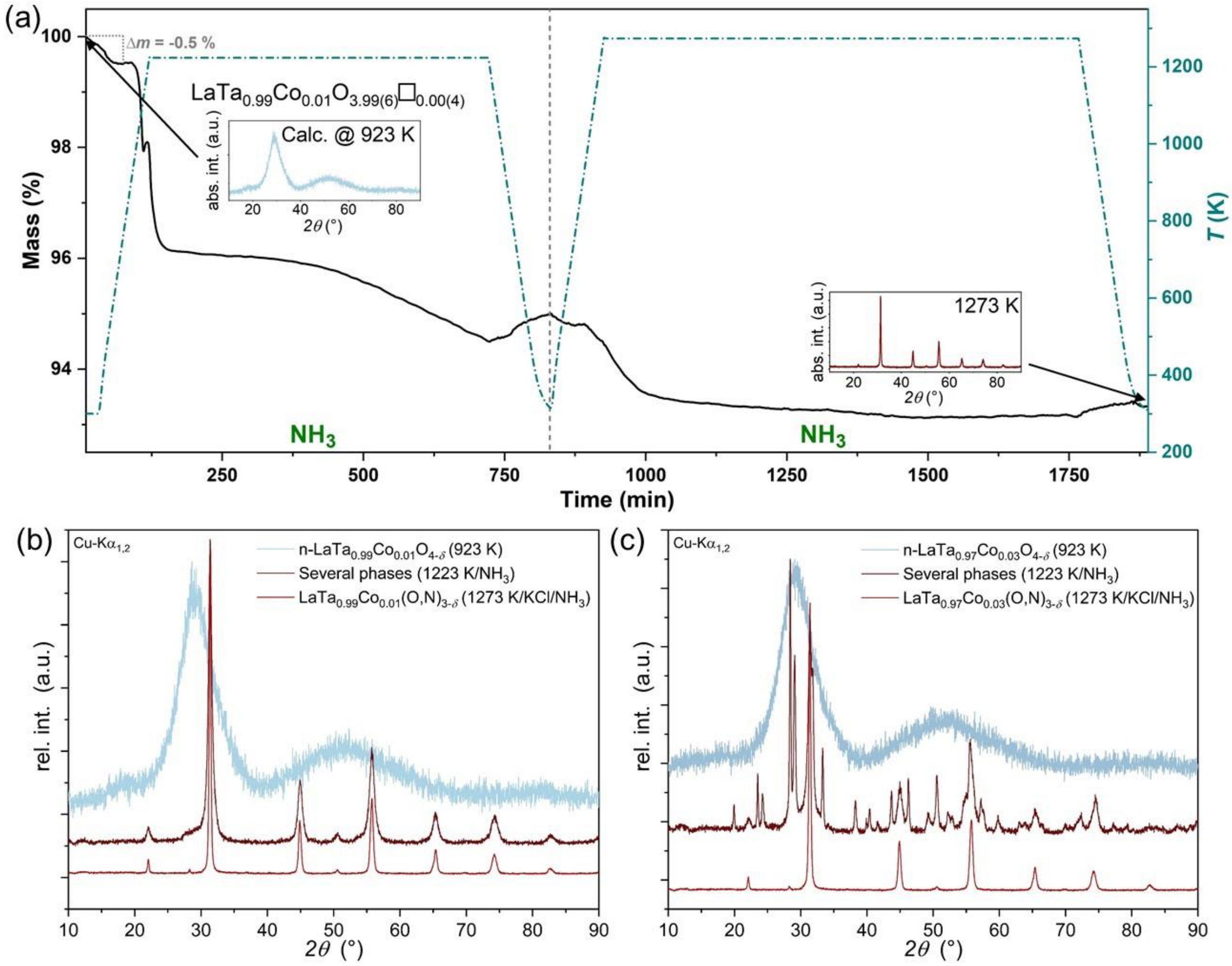

**Fig. S9.** *In situ* ammonolysis investigations *via* thermogravimetric analysis (TGA). (a) *In situ* ammonolysis of LTCO-1. For a better visibility of the oxygen vacancies the compound LTCO-1 is written in the figure as $LaTa_{0.99}Co_{0.01}O_{3.99(6)}\square_{0.00(4)}$. The square (□) represents the vacancies. The reaction steps observed by performing *in situ* ammonolysis (10 vol% Ar in $NH_3$) were similar to that of $LaTaO_2N$ [1]. However, one difference is observed for LTCO-1: After the first cycle the obtained oxynitride is according to PXRD not phase pure. The curve colors are reflecting the colors of the samples. (b)–(c) Powder diffraction patterns of the products of LTCON from the *ex situ* ammonolysis of the oxides (blue) to the oxynitrides.

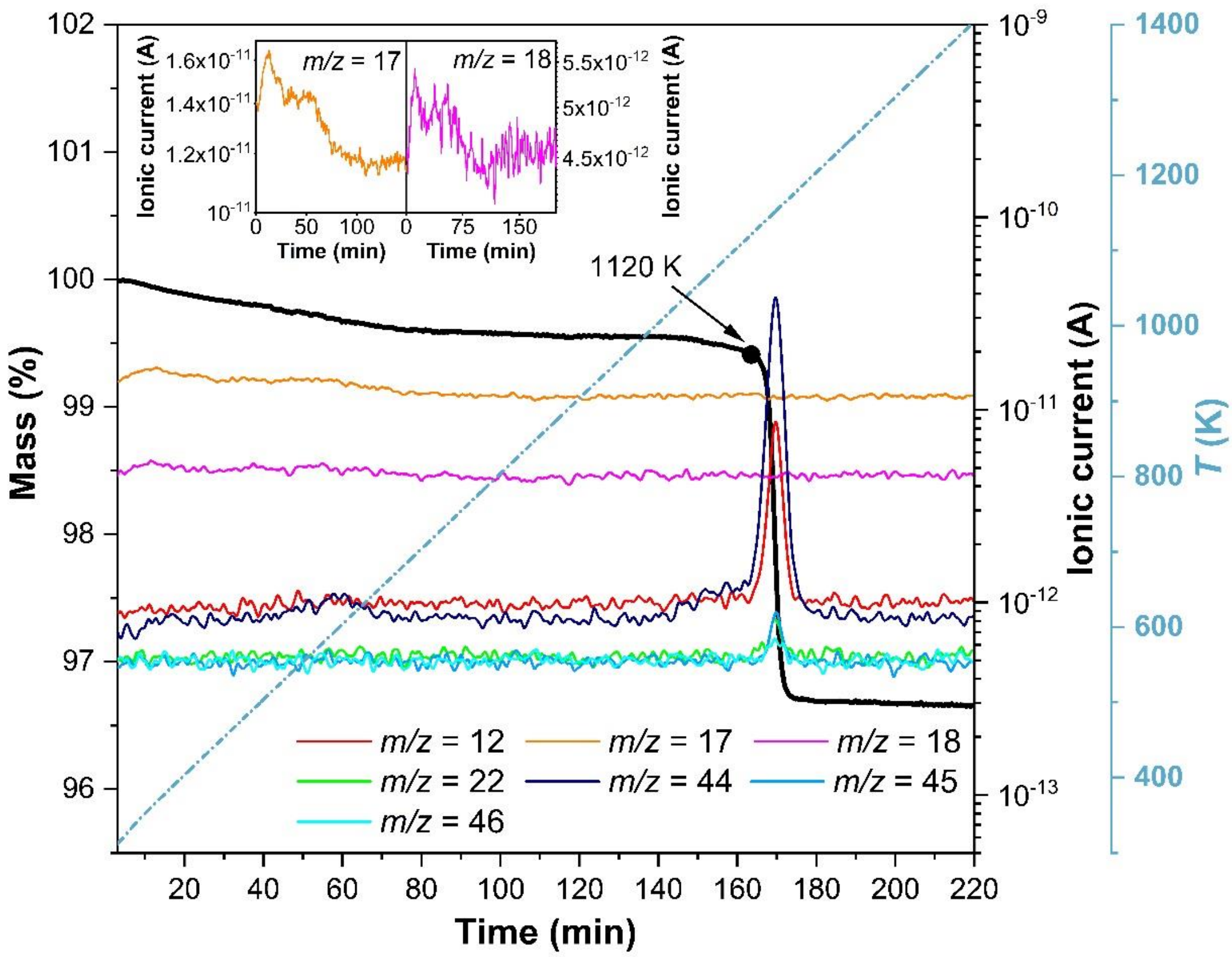

**Fig. S10.** Thermogravimetric analysis coupled with mass spectrometry (TGA-MS). TGA-MS experiment of LTCO-1 in synthetic air revealing adsorbed water and $CO_2$ pointing to organic residues on the sample's surface (5 K $min^{-1}$).

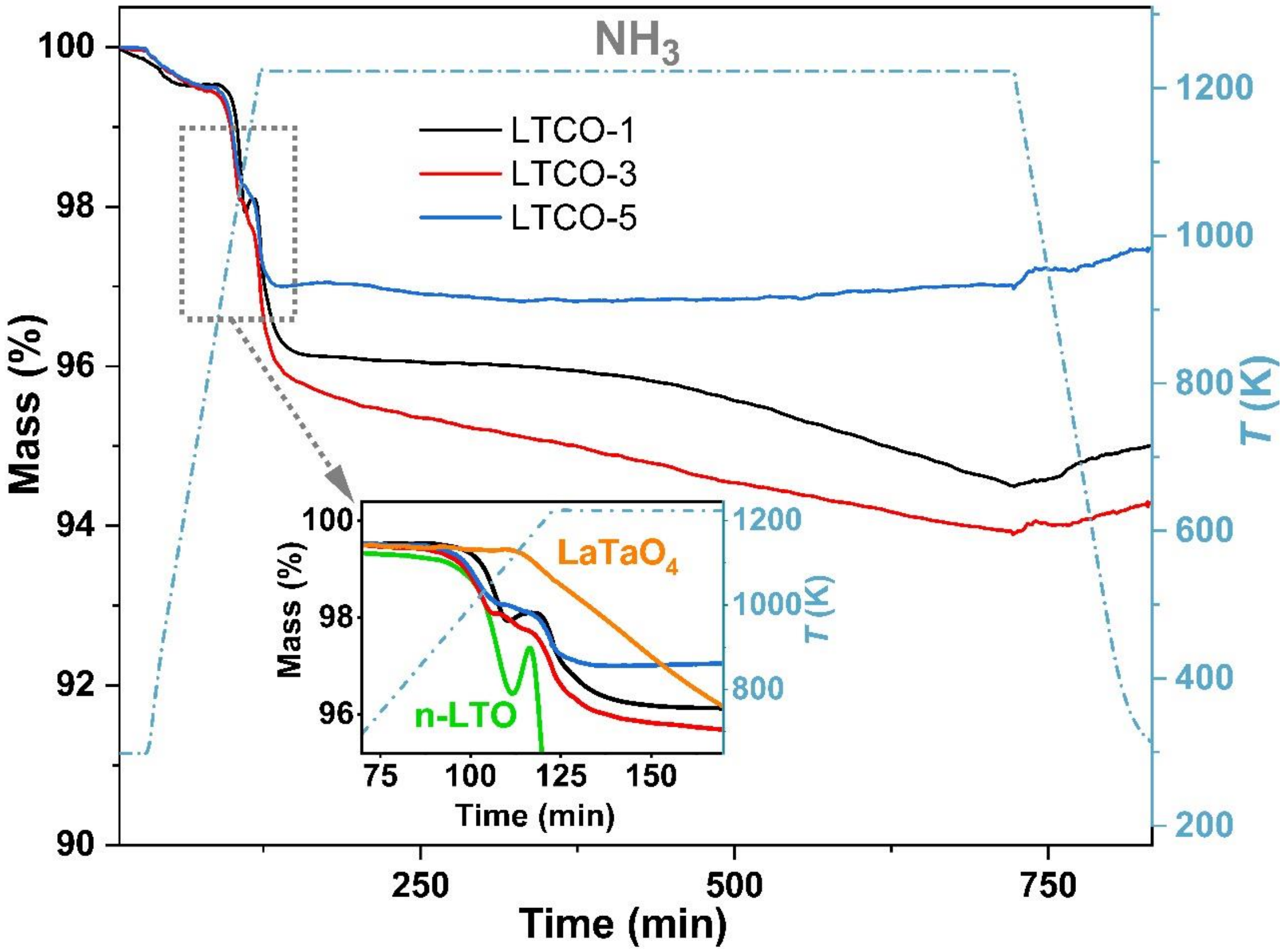

**Fig. S11.** *In situ* ammonolysis of the oxides. *In situ* ammonolysis (10 vol% Ar in $NH_3$) of LTCO. The reaction steps are similar to those of $LaTa^{IV}O_2N$ [1] on the first view. However, some differences were observed by the use of LTCO: The present oxygen vacancies lead to a continuous flattening of the local maximum between 1050 K and 1223 K (inset) and to an anionic ratio of nearly O : N = 1:2 in LTCON. The obtained curve progressions (black (LTCO-1), red (LTCO-3), and blue (LTCO-5)) lay in between the curve progressions of the unsubstituted nanocrystalline lanthanum tantalum oxide n-LTO (green) and microcrystalline $LaTaO_4$ (orange). The shown reference samples n-LTO and $LaTaO_4$ were synthesized according to ref. [1]. The dashed line represents the temperature.

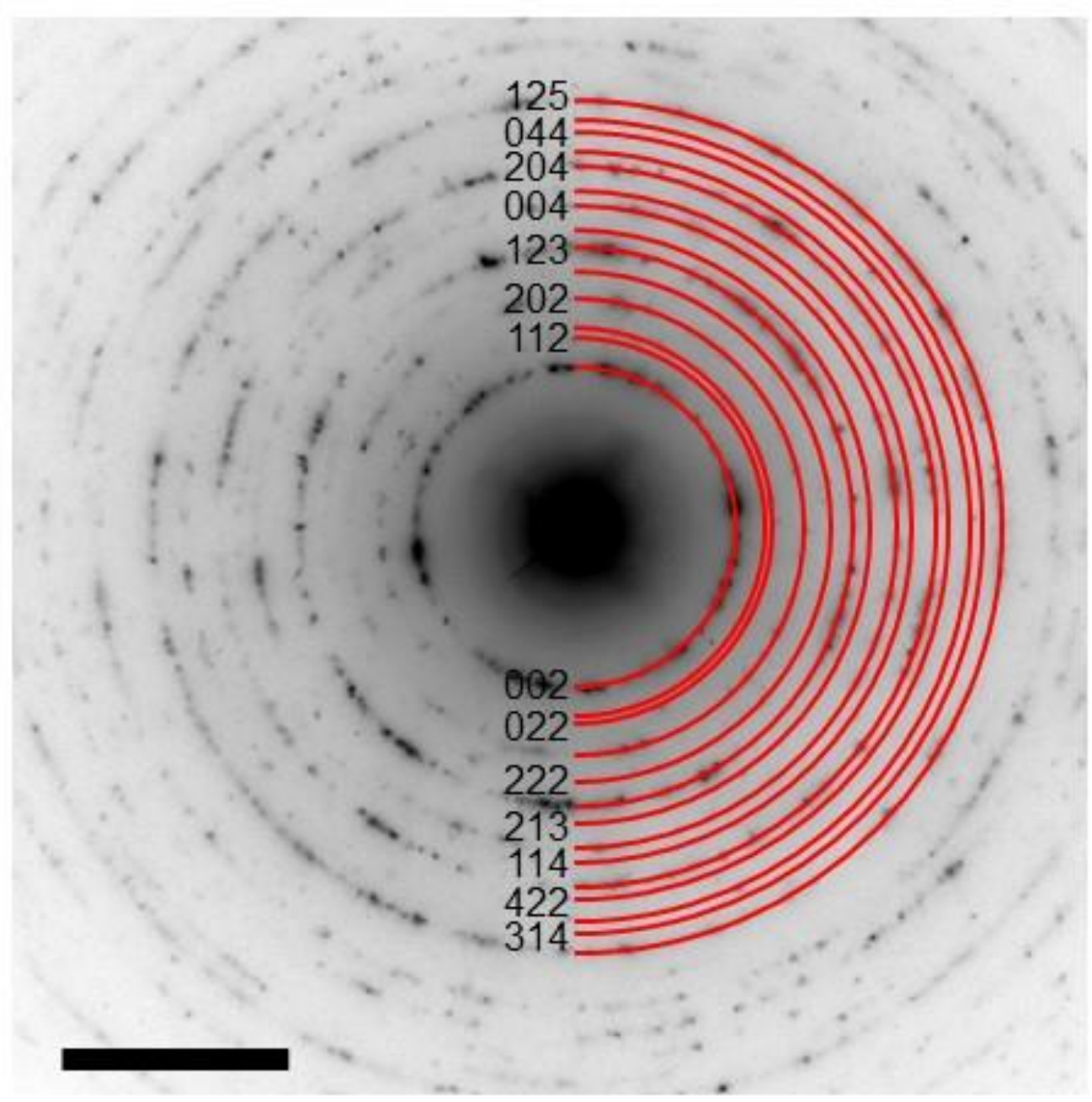

**Fig. S12.** Selected-area diffraction pattern of LTCON-5. Selected-area electron diffraction pattern and the ring annotations according to the *Imma* space group. Scale bar: 5 $nm^{-1}$.

**Tab. S4.** Compositions of LTCON after the third ammonolysis cycle. The compositions were determined by inductively coupled plasma optical emission spectroscopy (ICP-OES) and hot gas extraction (HGE).

| Compound | La[1] | Ta[1] | Co[1] | O[2] | N[2] | □[3] |
|---|---|---|---|---|---|---|
| **LTCON-1 (wt%)** | 39.3 ± 0.4 | 48.8 ± 0.5 | 0.20 ± 0.01 | 4.3 ± 0.1 | 7.65 ± 0.09 | |
| **LTCON-1** | 1.04 ± 0.01 | 0.98 ± 0.01 | 0.012 ± 0.007 | 0.96 ± 0.03 | 1.98 ± 0.02 | 0.05(5) |
| **LTCON-3 (wt%)** | 39 ± 0.4 | 47.9 ± 0.5 | 0.47 ± 0.03 | 4.38 ± 0.05 | 7.7 ± 0.1 | |
| **LTCON-3** | 1.03 ± 0.02 | 0.96 ± 0.01 | 0.029 ± 0.002 | 0.99 ± 0.01 | 1.97 ± 0.03 | 0.03(4) |
| **LTCON-5 (wt%)** | 39.4 ± 0.4 | 47.5 ± 0.5 | 0.80 ± 0.02 | 5.03 ± 0.06 | 7.21 ± 0.08 | |
| **LTCON-5** | 1.01 ± 0.01 | 0.94 ± 0.01 | 0.049 ± 0.001 | 1.13 ± 0.05 | 1.84 ± 0.02 | 0.03(3) |

[1] ICP-OES, [2] HGE, [3] Calculated vacancies obtained from HGE measurements

**Tab. S5.** Unit cell parameters of the oxynitrides. Unit cell parameters of single phase LTCON-1, LTCON-3, and LTCON-5 at 300 K from refinements of the PXRD data.

| Unit Cell Parameter | LTCON-1 | LTCON-3 | LTCON-5 |
|---|---|---|---|
| **Space group type** | *Imma* | *Imma* | *Imma* |
| $a$ (Å) | 5.7145(5) | 5.7147(2) | 5.7140(1) |
| $b$ (Å) | 8.0642(3) | 8.0647(1) | 8.0665(4) |
| $c$ (Å) | 5.7442(7) | 5.7447(4) | 5.7423(3) |
| $V_{cell}$ (Å$^3$) | 264.71(6) | 264.76(1) | 264.67(8) |
| $\rho$ (g·cm$^{-3}$) | 9.106 | 9.033 | 8.973 |
| **Phase fraction (wt%)** | 100 | 100 | 100 |
| $R_p$ (%) | 2.30 | 2.29 | 2.12 |
| $R_{wp}$ (%) | 3.13 | 3.17 | 2.94 |
| $\chi^2$ | 5.65 | 4.96 | 7.17 |
| $R_{Bragg}$ (%) | 1.06 | 2.40 | 1.48 |

**Tab. S6.** Refined atom positions of LTCON. The space group type *Imma* was determined by refinements of the PXRD data.

| Compound | Atom | Wyck. Symb. | *x* | *y* | *z* | $B_{iso}$ (Å$^2$) | sof[1] |
|---|---|---|---|---|---|---|---|
| | La | 4*e* | 0 | ¼ | 0.5[4] | 1.74(1) | 1[2] |
| | Ta | 4*a* | 0 | 0 | 0 | 1.49(7) | 0.99[2] |
| | Co | 4*a* | 0 | 0 | 0 | 1.49(7) | 0.01[2] |
| LTCON-1 | O(1) | 4*e* | 0 | ¼ | 0.1021(2) | 0.5[2] | 0.32564[3] |
| | N(1) | 4*e* | 0 | ¼ | 0.1021(2) | 0.5[2] | 0.68179[3] |
| | O(2) | 8*g* | ¼ | 0.9659(7) | ¼ | 0.5[2] | 0.32564[3] |
| | N(2) | 8*g* | ¼ | 0.9659(7) | ¼ | 0.5[2] | 0.68179[3] |
| | La | 4*e* | 0 | ¼ | 0.5[4] | 1.13(8) | 1[2] |
| | Ta | 4*a* | 0 | 0 | 0 | 0.71(8) | 0.97[2] |
| | Co | 4*a* | 0 | 0 | 0 | 0.71(8) | 0.03[2] |
| LTCON-3 | O(1) | 4*e* | 0 | ¼ | 0.0952(2) | 0.5[2] | 0.33920[3] |
| | N(1) | 4*e* | 0 | ¼ | 0.0952(2) | 0.5[2] | 0.65661[3] |
| | O(2) | 8*g* | ¼ | 0.9659(8) | ¼ | 0.5[2] | 0.33920[3] |
| | N(2) | 8*g* | ¼ | 0.9659(8) | ¼ | 0.5[2] | 0.65661[3] |
| | La | 4*e* | 0 | ¼ | 0.5[4] | 1.30(6) | 1[2] |
| | Ta | 4*a* | 0 | 0 | 0 | 0.75(2) | 0.95[2] |
| | Co | 4*a* | 0 | 0 | 0 | 0.75(2) | 0.05[2] |
| LTCON-5 | O(1) | 4*e* | 0 | ¼ | 0.0962(2) | 0.5[2] | 0.37524[3] |
| | N(1) | 4*e* | 0 | ¼ | 0.0962(2) | 0.5[2] | 0.61380[3] |
| | O(2) | 8*g* | ¼ | 0.9682(8) | ¼ | 0.5[2] | 0.37524[3] |
| | N(2) | 8*g* | ¼ | 0.9682(8) | ¼ | 0.5[2] | 0.61380[3] |

[1]site occupancy factor, [2]fixed, [3]fixed according to HGE results, [4]fixed according to Porter *et al*. [11]

**Tab. S7.** Distances in the $[(Ta,Co)(O,N)_6]^{z-}$ octahedra of LTCON. They were determined *via* Rietveld refinements of the respective PXRD data.

| Compound | LTCON-1 | LTCON-3 | LTCON-5 |
|---|---|---|---|
| $d_{(Ta,Co)-(O,N)1}$ (Å) | 2.047(2) | 2.040(3) | 2.0414(8) |
| $d_{(Ta,Co)-(O,N)2}$ (Å) | 2.047(2) | 2.040(3) | 2.0414(8) |
| $d_{(Ta,Co)-(O,N)3}$ (Å) | 2.047(2) | 2.040(3) | 2.0414(8) |
| $d_{(Ta,Co)-(O,N)4}$ (Å) | 2.047(2) | 2.040(3) | 2.0414(8) |
| $d_{(Ta,Co)-(O,N)5}$ (Å) | 2.057(3) | 2.076(8) | 2.091(2) |
| $d_{(Ta,Co)-(O,N)6}$ (Å) | 2.057(3) | 2.076(8) | 2.091(2) |
| $d_{(Ta,Co)-(O,N),average}$ (Å) | 2.050(6) | 2.052(5) | 2.058(1) |

**Tab. S8.** Angles in the $[(Ta,Co)(O,N)_6]^{z-}$ octahedra of LTCON. They were determined *via* Rietveld refinements of the respective PXRD data.

| Compound | LTCON-1 | LTCON-3 | LTCON-5 |
|---|---|---|---|
| $\measuredangle_{(Ta,Co)–X(1)–(Ta,Co)}$ (°) | 156.81(1) | 151.8(3) | 149.36(9) |
| $\measuredangle_{(Ta,Co)–X(2)–(Ta,Co)}$ (°) | 163.11(6) | 165.49(1) | 165.56(3) |
| $\measuredangle_{average}$ (°) | 159.96(4) | 158.7(2) | 157.4(7) |

**Tab. S9.** Unit cell parameters of LTCON-5 containing Co(O,N). Unit cell parameters of $LaTa_{0.95}Co_{0.05}(O,N)_{3-\delta}$ and Co(O,N) at 300 K from refinements of the PXRD data.

| Unit Cell Parameter | LTCON-5 | | Co(O,N) |
|---|---|---|---|
| **Space group type** | *Imma* | | $F\bar{4}3m$ |
| ***a*** **(Å)** | 5.7137(6) | | 4.12478[1] |
| ***b*** **(Å)** | 8.0625(6) | | |
| ***c*** **(Å)** | 5.7433(9) | | |
| $V_{cell}$ **(Å³)** | 264.58(4) | | 70.1784 |
| $\rho$ **(g·cm⁻³)** | 8.976 | | 6.998 |
| **Phase fraction (wt%)** | 99.4(8) | | 0.5(2) |
| $R_p$ **(%)** | | 2.15 | |
| $R_{wp}$ **(%)** | | 3.85 | |
| $\chi^2$ | | 3.70 | |
| $R_{Bragg}$ **(%)** | 2.33 | | 16.5 |

[1]fixed

**Tab. S10.** Atom positions of Co(O,N). The space group type $F\bar{4}3m$ was determined by refinements of the PXRD data. The atom positions were taken from Suzuki *et al*. [12], whereas the concentrations of Co, N, and O were determined by EDX.

| Atom | Wyck. Symb. | *x* | *y* | *z* | $B_{iso}$ (Å²) | sof[1] |
|---|---|---|---|---|---|---|
| Co | 4*a* | 0 | 0 | 0 | 1[2] | 1[2] |
| O | 4*c* | ¼ | ¼ | ¼ | 0.5[2] | 0.5[2] |
| N | 4*c* | ¼ | ¼ | ¼ | 0.5[2] | 0.5[2] |

[1]site occupancy factor, [2]fixed

**Tab. S11.** Unit cell parameters of single-phase LTCON-1 (from ND data). The V originates from the sample container.

| Temperature | Unit Cell Parameter | LTCON-1 | V |
|---|---|---|---|
| 300 K | Space group type | *Imma* | $Im\bar{3}m$ |
| | $a$ (Å) | 5.7112(4) | 3.0218(1) |
| | $b$ (Å) | 8.0574(5) | 3.0218(1) |
| | $c$ (Å) | 5.7458(4) | 3.0218(1) |
| | $V_{cell}$ (Å³) | 264.41(3) | 27.594(2) |
| | $\rho$ (g·cm$^{-3}$) | 9.117 | 6.132 |
| | Phase fraction (wt%) | 24.66(2) | 75.34(8) |
| | $R_p$ (%) | 2.87 | |
| | $R_{wp}$ (%) | 3.65 | |
| | $\chi^2$ | 1.31 | |
| | $R_{Bragg}$ (%) | 3.07 | 8.23 |
| 10 K | Space group type | *Imma* | $Im\bar{3}m$ |
| | $a$ (Å) | 5.7019(4) | 3.014(3) |
| | $b$ (Å) | 8.0471(6) | 3.014(3) |
| | $c$ (Å) | 5.7394(5) | 3.014(3) |
| | $V_{cell}$ (Å³) | 263.35(3) | 27.39(3) |
| | $\rho$ (g·cm$^{-3}$) | 9.154 | 6.178 |
| | Phase fraction (wt%) | 26.76(3) | 73.24(1) |
| | $R_p$ (%) | 3.24 | |
| | $R_{wp}$ (%) | 4.37 | |
| | $\chi^2$ | 1.65 | |
| | $R_{Bragg}$ (%) | 4.16 | 10.1 |

**Tab. S12.** Refined atom positions of LTCON-1 obtained from ND data. The space group type *Imma* was determined by refinements of the ND data.

| Temperature | Atom | Wyck. Symb. | $x$ | $y$ | $z$ | $B_{iso}$ (Å²) | sof[1] |
|---|---|---|---|---|---|---|---|
| 300 K | La | 4*e* | 0 | ¼ | 0.5[4] | 0.6[2] | 1[2] |
| | Ta | 4*a* | 0 | 0 | 0 | 0.3[2] | 0.99[2] |
| | Co | 4*a* | 0 | 0 | 0 | 0.3[2] | 0.01[2] |
| | O(1) | 4*e* | 0 | ¼ | 0.0686(4) | 0.771(6) | 0.32564[3] |
| | N(1) | 4*e* | 0 | ¼ | 0.0686(4) | 0.771(6) | 0.68179[3] |
| | O(2) | 8*g* | ¼ | 0.9633(5) | ¼ | 1.595(4) | 0.32564[3] |
| | N(2) | 8*g* | ¼ | 0.9633(5) | ¼ | 1.595(4) | 0.68179[3] |
| 10 K | La | 4*e* | 0 | ¼ | 0.5[4] | 0.6[2] | 1[2] |
| | Ta | 4*a* | 0 | 0 | 0 | 0.3[2] | 0.99[2] |
| | Co | 4*a* | 0 | 0 | 0 | 0.3[2] | 0.01[2] |
| | O(1) | 4*e* | 0 | ¼ | 0.0691(6) | 0.873(8) | 0.32564[3] |
| | N(1) | 4*e* | 0 | ¼ | 0.0691(6) | 0.873(8) | 0.68179[3] |
| | O(2) | 8*g* | ¼ | 0.9622(3) | ¼ | 1.595(5) | 0.32564[3] |
| | N(2) | 8*g* | ¼ | 0.9622(3) | ¼ | 1.595(5) | 0.68179[3] |

[1]site occupancy factor, [2]fixed, [3]fixed according to HGE results, [4]fixed according to Porter *et al*. [11]

**Tab. S13.** Distances in the $[(Ta,Co)(O,N)_6]^{z-}$ octahedron of LTCON-1. They were determined *via* Rietveld refinements of the respective ND data.

| Compound | LTCON-1 (300 K) | LTCON-1 (10 K) |
|---|---|---|
| $d_{(Ta,Co)\text{-}(O,N)1}$ (Å) | 2.0468(3) | 2.0453(4) |
| $d_{(Ta,Co)\text{-}(O,N)2}$ (Å) | 2.0468(3) | 2.0453(4) |
| $d_{(Ta,Co)\text{-}(O,N)3}$ (Å) | 2.0468(3) | 2.0453(4) |
| $d_{(Ta,Co)\text{-}(O,N)4}$ (Å) | 2.0468(3) | 2.0453(4) |
| $d_{(Ta,Co)\text{-}(O,N)5}$ (Å) | 2.0526(6) | 2.0505(7) |
| $d_{(Ta,Co)\text{-}(O,N)6}$ (Å) | 2.0526(6) | 2.0505(7) |
| $d_{(Ta,Co)\text{-}(O,N),average}$ (Å) | 2.0487(5) | 2.0470(8) |

**Tab. S14.** Angles in the $[(Ta,Co)(O,N)_6]^{z-}$ octahedron of LTCON-1.They were determined *via* Rietveld refinements of the respective ND data.

| Compound | LTCON-1 (300 K) | LTCON-1 (10 K) |
|---|---|---|
| $\measuredangle_{(Ta,Co)-X(1)-(Ta,Co)}$ (°) | 157.86(3) | 157.70(3) |
| $\measuredangle_{(Ta,Co)-X(2)-(Ta,Co)}$ (°) | 163.39(2) | 162.89(4) |
| $\measuredangle_{average}$ (°) | 160.62(7) | 160.29(8) |

**Tab. S15.** Unit cell parameters of single-phase LTCON-3 (from ND data). The V originates from the sample container.

| Temperature | Unit Cell Parameter | LTCON-3 | V |
|---|---|---|---|
| | Space group type | *Imma* | $Im\bar{3}m$ |
| | $a$ (Å) | 5.7134(4) | 3.026(1) |
| | $b$ (Å) | 8.0609(5) | 3.026(1) |
| | $c$ (Å) | 5.7484(4) | 3.026(1) |
| | $V_{cell}$ (Å³) | 264.74(3) | 27.71(2) |
| 300 K | $\rho$ (g·cm$^{-3}$) | 9.034 | 6.106 |
| | Phase fraction (wt%) | 25.05(2) | 74.95(8) |
| | $R_p$ (%) | 2.81 | |
| | $R_{wp}$ (%) | 3.57 | |
| | $\chi^2$ | 1.27 | |
| | $R_{Bragg}$ (%) | 2.80 | 10.9 |
| | Space group type | *Imma* | $Im\bar{3}m$ |
| | $a$ (Å) | 5.7057(5) | 3.020(2) |
| | $b$ (Å) | 8.0519(7) | 3.020(2) |
| | $c$ (Å) | 5.7409(5) | 3.020(2) |
| | $V_{cell}$ (Å³) | 263.75(4) | 27.53(3) |
| 10 K | $\rho$ (g·cm$^{-3}$) | 9.007 | 6.143 |
| | Phase fraction (wt%) | 25.03(2) | 74.97(1) |
| | $R_p$ (%) | 3.21 | |
| | $R_{wp}$ (%) | 4.33 | |
| | $\chi^2$ | 1.58 | |
| | $R_{Bragg}$ (%) | 3.37 | 8.11 |

**Tab. S16.** Refined atom positions of LTCON-3 obtained from ND data. The space group type *Imma* is determined by refinements of the ND data.

| Temperature | Atom | Wyck. Symb. | *x* | *y* | *z* | $B_{iso}$ (Å²) | sof[1] |
|---|---|---|---|---|---|---|---|
| 300 K | La | 4*e* | 0 | ¼ | 0.5[4] | 0.6[2] | 1[2] |
| | Ta | 4*a* | 0 | 0 | 0 | 0.3[2] | 0.97[2] |
| | Co | 4*a* | 0 | 0 | 0 | 0.3[2] | 0.03[2] |
| | O(1) | 4*e* | 0 | ¼ | 0.0690(4) | 0.613(6) | 0.33920[3] |
| | N(1) | 4*e* | 0 | ¼ | 0.0690(4) | 0.613(6) | 0.65661[3] |
| | O(2) | 8*g* | ¼ | 0.9637(2) | ¼ | 1.670(4) | 0.33920[3] |
| | N(2) | 8*g* | ¼ | 0.9637(2) | ¼ | 1.670(4) | 0.65661[3] |
| 10 K | La | 4*e* | 0 | ¼ | 0.5[4] | 0.6[2] | 1[2] |
| | Ta | 4*a* | 0 | 0 | 0 | 0.3[2] | 0.97[2] |
| | Co | 4*a* | 0 | 0 | 0 | 0.3[2] | 0.03[2] |
| | O(1) | 4*e* | 0 | ¼ | 0.0703(5) | 0.570(7) | 0.33920[3] |
| | N(1) | 4*e* | 0 | ¼ | 0.0703(5) | 0.570(7) | 0.65661[3] |
| | O(2) | 8*g* | ¼ | 0.9626(3) | ¼ | 1.738(6) | 0.33920[3] |
| | N(2) | 8*g* | ¼ | 0.9626(3) | ¼ | 1.738(6) | 0.65661[3] |

[1]site occupancy factor, [2]fixed, [3]fixed according to HGE results, [4]fixed according to Porter *et al*. [11]

**Tab. S17.** Distances in the $[(Ta,Co)(O,N)_6]^{z-}$ octahedron of LTCON-3. They were determined *via* Rietveld refinements of the respective ND data.

| Compound | LTCON-3 (300 K) | LTCON-3 (10 K) |
|---|---|---|
| $d_{\text{(Ta,Co)-(O,N)1}}$ (Å) | 2.0472(3) | 2.0458(4) |
| $d_{\text{(Ta,Co)-(O,N)2}}$ (Å) | 2.0472(3) | 2.0458(4) |
| $d_{\text{(Ta,Co)-(O,N)3}}$ (Å) | 2.0472(3) | 2.0458(4) |
| $d_{\text{(Ta,Co)-(O,N)4}}$ (Å) | 2.0472(3) | 2.0458(4) |
| $d_{\text{(Ta,Co)-(O,N)5}}$ (Å) | 2.0539(5) | 2.0530(6) |
| $d_{\text{(Ta,Co)-(O,N)6}}$ (Å) | 2.0539(5) | 2.0530(6) |
| $d_{\text{(Ta,Co)-(O,N),average}}$ (Å) | 2.0494(7) | 2.0482(5) |

**Tab. S18.** Angles in the $[(Ta,Co)(O,N)_6]^{z-}$ octahedron of LTCON-3. They were determined *via* Rietveld refinements of the respective ND data.

| Compound | LTCON-3 (300 K) | LTCON-3 (10 K) |
|---|---|---|
| $\measuredangle_{\text{(Ta,Co)–}X\text{(1)–(Ta,Co)}}$ (°) | 157.73(2) | 157.33(3) |
| $\measuredangle_{\text{(Ta,Co)–}X\text{(2)–(Ta,Co)}}$ (°) | 163.57(2) | 163.07(2) |
| $\measuredangle_{\text{average}}$ (°) | 160.65(2) | 160.20(3) |

**Tab. S19.** Unit cell parameters of single-phase LTCON-5 (from ND data). The V originates from the sample container.

| Temperature | Unit Cell Parameter | LTCON-5 | V |
|---|---|---|---|
| 300 K | Space group type | *Imma* | $Im\bar{3}m$ |
| | $a$ (Å) | 5.7108(3) | 3.022(1) |
| | $b$ (Å) | 8.0565(5) | 3.022(1) |
| | $c$ (Å) | 5.7465(4) | 3.022(1) |
| | $V_{cell}$ (Å³) | 264.39(3) | 27.58(3) |
| | $\rho$ (g·cm$^{-3}$) | 8.988 | 6.133 |
| | Phase fraction (wt%) | 25.45(2) | 74.55(8) |
| | $R_p$ (%) | 2.89 | |
| | $R_{wp}$ (%) | 3.74 | |
| | $\chi^2$ | 1.36 | |
| | $R_{Bragg}$ (%) | 3.04 | 10.6 |
| 10 K | Space group type | *Imma* | $Im\bar{3}m$ |
| | $a$ (Å) | 5.7017(5) | 3.017(3) |
| | $b$ (Å) | 8.0460(7) | 3.017(3) |
| | $c$ (Å) | 5.7390(6) | 3.017(3) |
| | $V_{cell}$ (Å³) | 263.28(4) | 27.45(5) |
| | $\rho$ (g·cm$^{-3}$) | 9.027 | 6.167 |
| | Phase fraction (wt%) | 33.2(7) | 66.7(3) |
| | $R_p$ (%) | 3.90 | |
| | $R_{wp}$ (%) | 5.38 | |
| | $\chi^2$ | 2.15 | |
| | $R_{Bragg}$ (%) | 9.55 | 24.2 |

**Tab. S20.** Refined atom positions of LTCON-5 obtained from ND data. The space group type *Imma* is determined by refinements of the ND data.

| Temperature | Atom | Wyck. Symb. | $x$ | $y$ | $z$ | $B_{iso}$ (Å²) | sof[1] |
|---|---|---|---|---|---|---|---|
| 300 K | La | 4*e* | 0 | ¼ | 0.5[4] | 0.6[2] | 1[2] |
| | Ta | 4*a* | 0 | 0 | 0 | 0.3[2] | 0.95[2] |
| | Co | 4*a* | 0 | 0 | 0 | 0.3[2] | 0.05[2] |
| | O(1) | 4*e* | 0 | ¼ | 0.0705(5) | 0.539(6) | 0.37524[3] |
| | N(1) | 4*e* | 0 | ¼ | 0.0705(5) | 0.539(6) | 0.61380[3] |
| | O(2) | 8*g* | ¼ | 0.9633(2) | ¼ | 1.451(5) | 0.37524[3] |
| | N(2) | 8*g* | ¼ | 0.9633(2) | ¼ | 1.451(5) | 0.61380[3] |
| 10 K | La | 4*e* | 0 | ¼ | 0.5[4] | 0.6[2] | 1[2] |
| | Ta | 4*a* | 0 | 0 | 0 | 0.3[2] | 0.95[2] |
| | Co | 4*a* | 0 | 0 | 0 | 0.3[2] | 0.05[2] |
| | O(1) | 4*e* | 0 | ¼ | 0.0696(8) | 1.044(9) | 0.37524[3] |
| | N(1) | 4*e* | 0 | ¼ | 0.0696(8) | 1.044(9) | 0.61380[3] |
| | O(2) | 8*g* | ¼ | 0.9620(4) | ¼ | 1.688(7) | 0.37524[3] |
| | N(2) | 8*g* | ¼ | 0.9620(4) | ¼ | 1.688(7) | 0.61380[3] |

[1]site occupancy factor, [2]fixed, [3]fixed according to HGE results, [4]fixed according to Porter *et al*. [11]

**Tab. S21.** Distances in the $[(Ta,Co)(O,N)_6]^{z-}$ octahedron of LTCON-5. They were determined *via* Rietveld refinements of the respective ND data.

| Compound | LTCON-5 (300 K) | LTCON-5 (10 K) |
|---|---|---|
| $d_{(Ta,Co)\text{-}(O,N)1}$ (Å) | 2.0469(2) | 2.0455(5) |
| $d_{(Ta,Co)\text{-}(O,N)2}$ (Å) | 2.0469(2) | 2.0455(5) |
| $d_{(Ta,Co)\text{-}(O,N)3}$ (Å) | 2.0469(2) | 2.0455(5) |
| $d_{(Ta,Co)\text{-}(O,N)4}$ (Å) | 2.0469(2) | 2.0455(5) |
| $d_{(Ta,Co)\text{-}(O,N)5}$ (Å) | 2.0545(6) | 2.0509(8) |
| $d_{(Ta,Co)\text{-}(O,N)6}$ (Å) | 2.0545(6) | 2.0509(8) |
| $d_{(Ta,Co)\text{-}(O,N),average}$ (Å) | 2.0494(7) | 2.0473(6) |

**Tab. S22.** Angles in the $[(Ta,Co)(O,N)_6]^{z-}$ octahedron of LTCON-5. They were determined *via* Rietveld refinements of the respective ND data.

| Compound | LTCON-5 (300 K) | LTCON-5 (10 K) |
|---|---|---|
| $\measuredangle_{(Ta,Co)–X1–(Ta,Co)}$ (°) | 157.25(3) | 157.54(4) |
| $\measuredangle_{(Ta,Co)–X2–(Ta,Co)}$ (°) | 163.38(9) | 162.71(3) |
| $\measuredangle_{average}$ (°) | 160.32(6) | 160.12(9) |

**Supplemental Material Sec. III: Critical assessment of the Co ion concentration**

We found that parameters such as the exact stoichiometric weighed ratio of the cations, and the amount of weighed oxide precursors prior to ammonolysis are crucial to get single-phase oxynitrides. By weighing 350 mg of LTCO containing the exact cationic stoichiometric amount single phase LTCON can be synthesized. The homogeneous distribution of the Co ions can be seen *via* transmission electron microscopy (TEM) and scanning electron microscopy (SEM) studies (Figs. S13–S15). In contrast, by weighing 350 mg of LTCO-5 with a slightly higher Co concentration ($x = 0.0517$), Co-rich nanoparticles with a diameter of 40 nm < $d$ < 80 nm (HR-TEM investigations, Figs. 2 and 3) can be obtained. This is because of the non-stoichiometric weighed amount of the cations in the samples (too much Co in comparison to La and Ta). The composition of the Co-rich nanoparticles determined by EDX is Co(O,N) (Co : O : N = 50 at% : 25 at% : 25 at%). A phase with a similar composition ($CoO_{0.74}N_{0.24}$) was already synthesized elsewhere [13].

Both observations – single-phase oxynitrides and an oxynitride containing a Co-rich phase – were investigated further by scanning electron microscopy (SEM), transmission electron microscopy (TEM), and high-resolution transmission electron microscopy (HR-TEM) (Figs. 2, 3 and Figs. S13–S15). All three techniques were coupled with energy-dispersive X-ray spectroscopy (EDX). Because objective polepieces can contain cobalt a false Co concentration in the particles can be detected. Therefore, two different TEM devices – JEOL ARM200F and a Philips CM-200 FEG – were selected. The latter device contains a pure iron polepiece, whereas the polepiece of the JEOL ARM200F contains Co. By measuring a Co-free reference sample ($LaTaON_2$), using the JEOL ARM200F, the Co : Fe ratio was determined to Co : Fe = 0.7 (Tab. S23). This factor was used to discriminate the Co in the specimen from the Co in the polepiece using Supplementary Equation 1:

$$\frac{I_{\mathrm{Co}}}{I_{\mathrm{Fe}}} = \frac{I_{\mathrm{Co,\ polepiece}} + I_{\mathrm{Co,\ sample}}}{I_{\mathrm{Fe,\ polepiece}}} = \frac{I_{\mathrm{Co,\ polepiece}}}{I_{\mathrm{Fe,\ polepiece}}} + \frac{I_{\mathrm{Co,\ sample}}}{I_{\mathrm{Fe,\ polepiece}}} = \frac{I_{\mathrm{Co,\ sample}}}{I_{\mathrm{Fe,\ polepiece}}} + 0.7 \qquad (1)$$

This ratio is considered by determining the Co ion concentration in the Co-containing LTCON samples. The obtained results of the Co ion concentrations are fitting well with the expected values (Tab. S23).

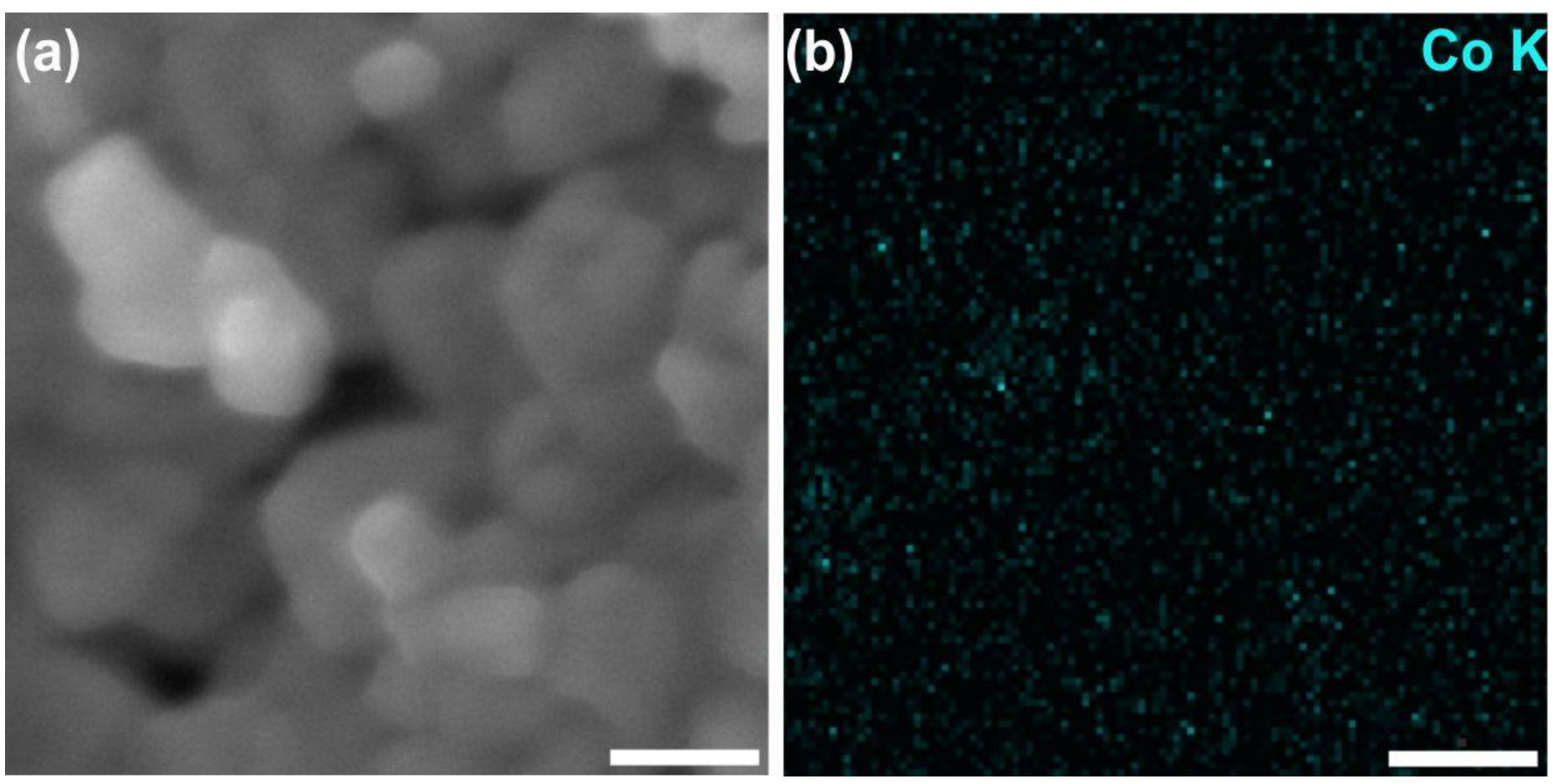

**Fig. S13.** SEM image and Co-*K* EDX map of LTCON-5 particles. (a) SEM image and (b) EDX map of agglomerated LTCON-5 nanoparticles showing the homogeneous distribution of Co ions in the particles. Scale bars: 200 nm.

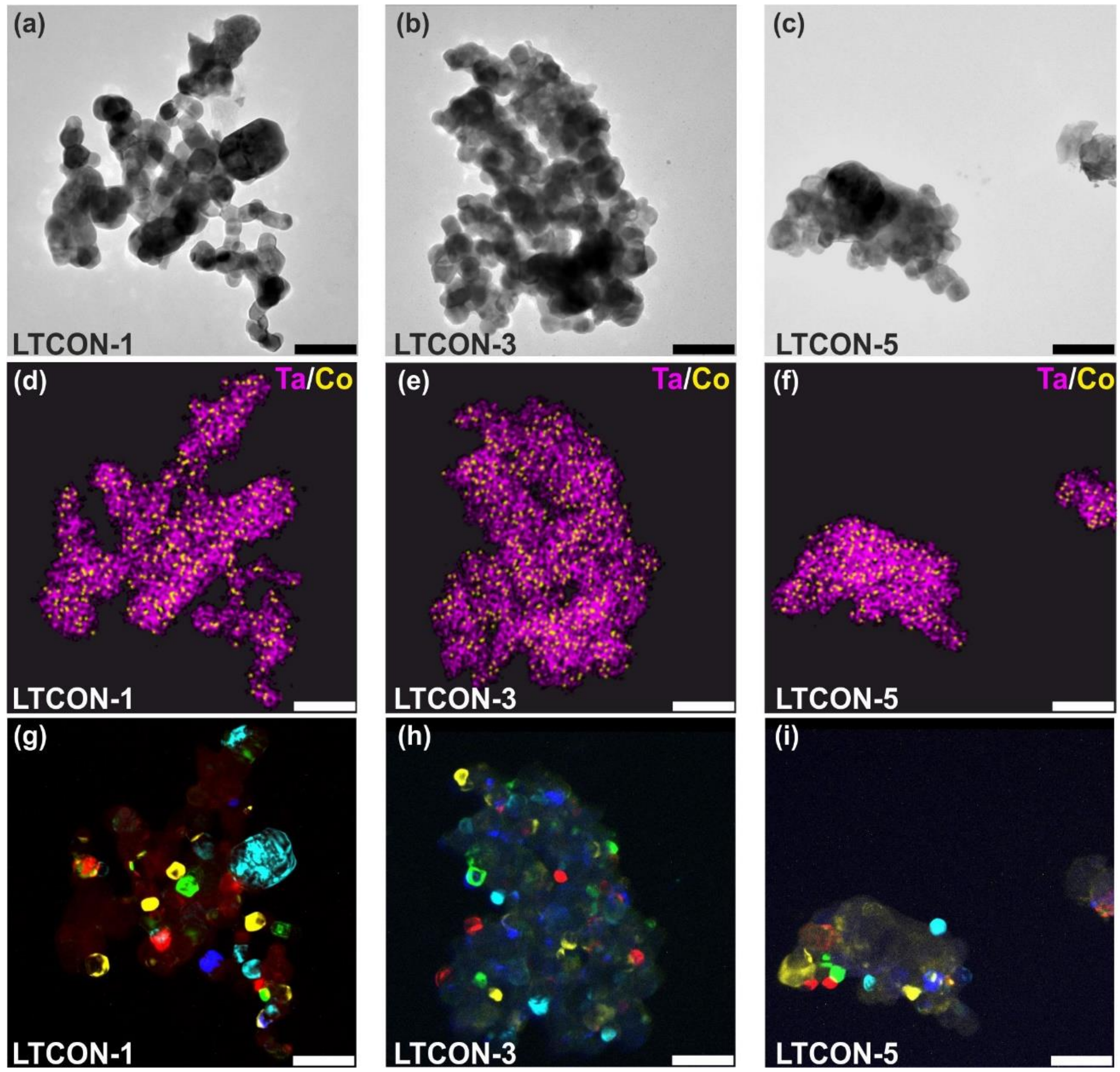

**Fig. S14.** TEM investigation of LTCON. (a)–(c) TEM images of particles of LTCON-1, LTCON-3, and LTCON-5 and (d)–(f) EDX maps of LTCON-1, LTCON-3, and LTCON-5 showing the homogeneous distribution of Co (yellow) and Ta (pink) in the particles. Elemental Co clusters or Co-rich secondary particles with a size larger than $d$ = 4 nm can be excluded because of the pixel size. (g)–(i) The dark-field TEM images show in different colors the different crystal orientations of the particles. All scale bars: 100 nm.

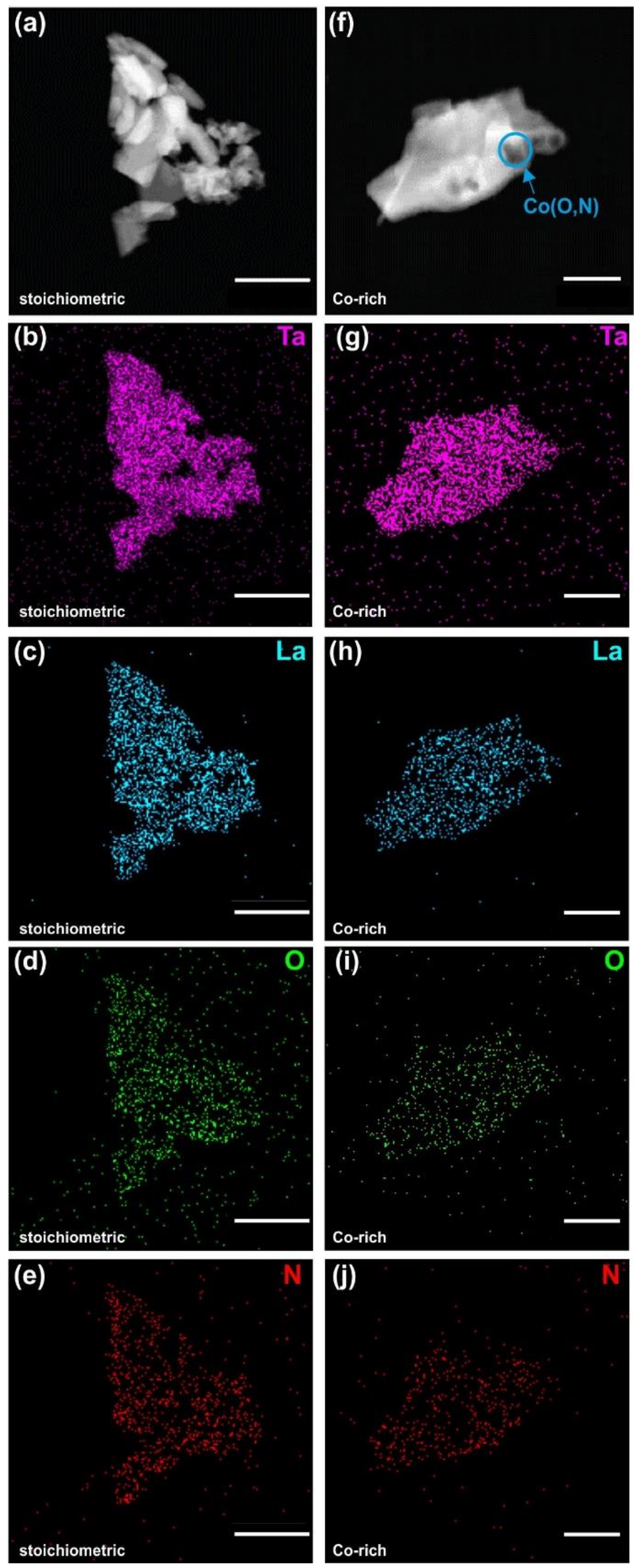

**Fig. S15.** HAADF/EDX investigation of LTCON-5. (a)–(e) Respective high-angle annular dark-field (HAADF) image and EDX maps of Ta, La, O, and N of single-phase (stoichiometric) LTCON-5 nanoparticles showing the homogeneous distribution of the elements. Scale bars: 500 nm. (f)–(j) Respective HAADF image and EDX maps of Ta, La, O, and N of LTCON-5 nanoparticles containing Co(O,N) nanoparticles showing the homogeneous distribution of Ta, La, O, and N. Scale bars: 250 nm.

**Tab. S23.** Measured Co and Fe intensities and determination of the Co concentration. The determination of the Co concentration is determined as described in the supplementary text for the HR-TEM JEOL microscope with a Co/Fe polepiece (cf. Supplementary Equation 1). Different lines represent measurements performed at different particles. The intensities are given in netcounts.

| **Sample** | Co (counts) | Fe (counts) | Co : Fe | Co : Fe (corrected) | $c_{Co}$ (at%) | $c_{Co,corrected}$ (at%) | $c_{Co, expected}$ (at%) | $c_{Co, determined}$ (at%) |
|---|---|---|---|---|---|---|---|---|
| | 3468 | 1780 | 1.95 | 1.25 | 0.76 | 0.53 | | |
| | 3121 | 1362 | 2.29 | 1.59 | 0.9 | 0.63 | | |
| **LTCON-3** | 6023 | 1914 | 3.15 | 2.45 | 0.78 | 0.54 | 0.6 | 0.55 |
| | 21921 | 7037 | 3.12 | 2.42 | 0.82 | 0.57 | | |
| | 6963 | 2598 | 2.68 | 1.98 | 0.69 | 0.48 | | |
| | 1289 | 1570 | 0.82 | 0.12 | 1.26 | 0.88 | | |
| | 1547 | 1428 | 1.08 | 0.38 | 1.48 | 1.03 | | |
| | 1332 | 629 | 2.12 | 1.42 | 1.19 | 0.83 | | |
| **LTCON-5** | 1508 | 607 | 2.48 | 1.79 | 1.24 | 0.87 | 1 | 0.83 |
| | 6864 | 2872 | 2.39 | 1.69 | 1.1 | 0.77 | | |
| | 2368 | 961 | 2.46 | 1.77 | 1.29 | 0.90 | | |
| | 2047 | 720 | 2.84 | 2.14 | 1.12 | 0.78 | | |
| | 2054 | 2756 | | | | | | |
| **$LaTaON_2$** | 1790 | 2177 | 0.7 | | | | 0 | 0 |
| | 1483 | 1941 | | | | | | |
| | 131 | 283 | | | | | | |

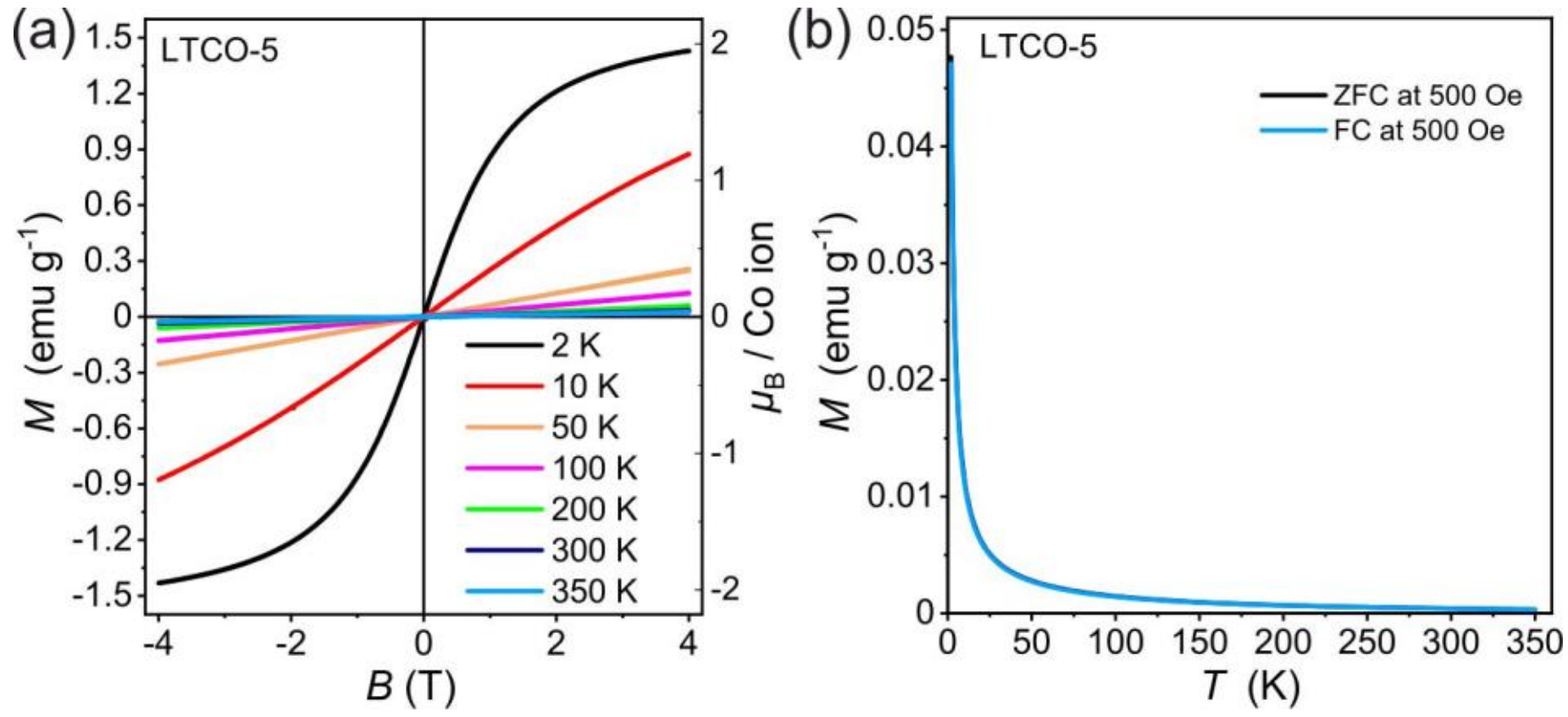

**Fig. S16.** SQUID-based magnetic measurements of LTCO-5. (a) Field-dependent magnetic measurement of LTCO-5. (b) Temperature-dependent magnetic measurement of LTCO-5. Both measurements reveal for the precursors simple paramagnetism over the entire measurement range.

**Supplemental Material Sec. IV: Comparison between raw and processed SQUID data**

In comparison to many earlier studies [14], in particular on Co-doped ZnO, the observed SQUID signal for LTCON is not dominated by a large diamagnetic background and a small ferromagnetic-like superimposed signal. For demonstration, we present in Fig. S17 a direct comparison between the raw (measured) and processed SQUID data of all LTCON samples. It is obvious that the relative amount of diamagnetic background increases with decreasing Co ion concentrations. Since for LTCON-3 (0.6 at% Co) almost no diamagnetism is present, the diamagnetic background for LTCON-1 (0.2 at% Co) is more prominent because of the lower Co ion concentration and with it the smaller ferromagnetic contribution. Nevertheless, in contrast to many other studies published on diluted magnetic semiconductors (DMS) (see referenced in the main text) the ferromagnetic contribution is quite large with 3–5 orders of magnitude above the detection limit of the SQUID device.

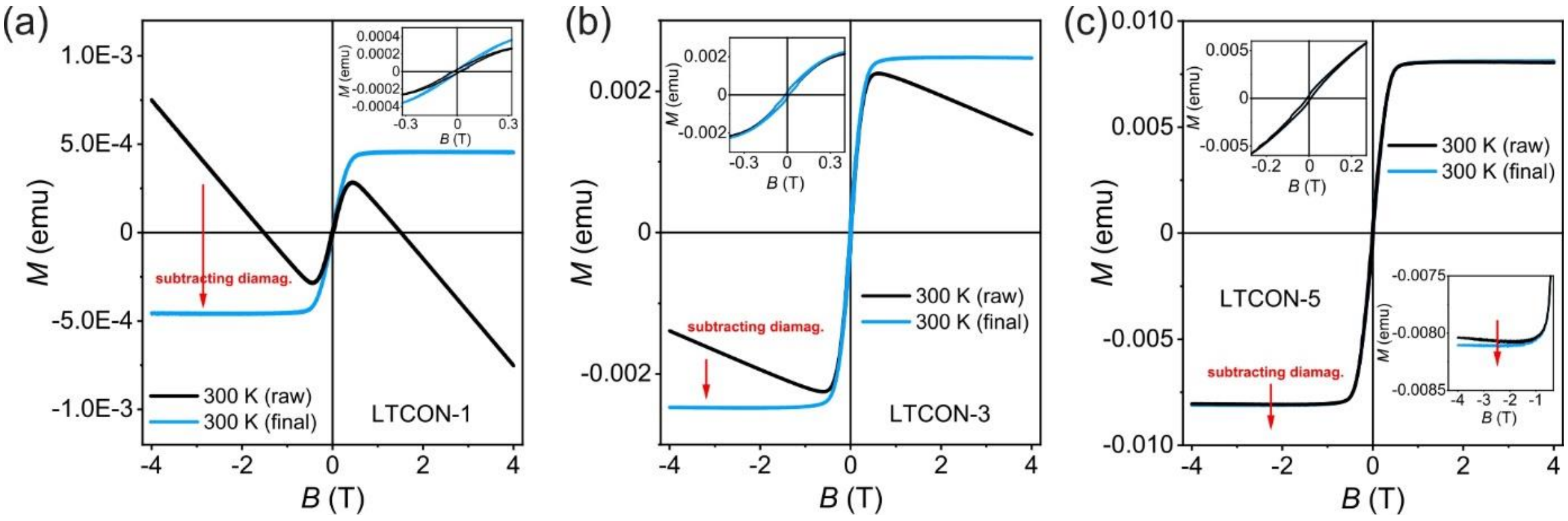

**Fig. S17.** Raw and diamagnetic background-corrected SQUID data of LTCON. (a)–(c) SQUID data (raw and processed) for all Co ion concentrations. The diamagnetic background was determined as the negative slope of the raw data.

**Supplemental Material Sec. V: Exclusion of elemental Co clusters and secondary Co-rich phases**

To compare the effect of Co-rich nanoparticles in LTCON-5 and single-phase LTCON-5 SQUID-based magnetometry was applied (Fig. 4d). This was done in order to identify possible Co-rich phases and to further proof if the high-temperature ferromagnetism is truly originating from LTCON. In Fig. 4d, normalized 300 K magnetization curves of the stoichiometric and the Co-rich (non-stoichiometric) samples are presented. In Fig. 4d both curves look quite similar in shape. By magnifying the low-field region - as shown in Fig. 4d - a small difference is observable where the Co-rich sample reveals an additional contribution, with a higher initial slope, as indicated by the blue curve. The purple arrows indicate a difference of about 7 % of the saturation value. By plotting the difference between both samples and normalize their saturation to unity, the additional phase as shown in by the orange curve can be identified. This contribution has a much higher slope compared to the main curves, which we interpret as a part of Co(O,N). Nevertheless, it is obvious that even for the Co-rich sample the dominating part is related to the main-phase magnetism, also consistent to our very rough bulk estimates as discussed above. Those larger nanoparticles reveal a much steeper hysteresis curve at 300 K as expected.

In Fig. S18a) calculated superparamagnetic Langevin-Function curves for an approximate 4 nm sized elemental Co cluster together with the measured room temperature (300 K) $M(H)$ curve of the LTCON-5 are presented (open circles). 3047 Co atoms per 4 nm cluster were assumed. We have chosen the size of 4 nm, because the 300 K behavior is most similar to our experimental results and the lowest resolution which was used for TEM experiments (Fig. S14) was 4 nm. As no temperature-dependent significant shape variations were observed for LTCON-5, only the 300 K magnetization curve (Fig. 4c) was used for comparison. The shape of the measured 300 K $M(H)$ curve of LTCON-5 is by any means not consistent with all of the calculated temperature-dependent Langevin-functions [15]. In contradiction to the superparamagnetic model, the LTCON samples do not reveal any significant shape variations from 2 K up to 350 K. For example, Co-hcp as secondary phase in Co-doped ZnO shows these shape variations [14,16]. Therefore, the presence of superparamagnetic clusters made of any ferromagnetic material can be excluded as an explanation to our high-temperature ferromagnetism above room temperature. Especially, smaller clusters reveal an even stronger temperature dependence [17]. This can be seen nicely in Fig. S18b, where the size-dependence is presented at 300 K. Again, no simulated curve shape is consistent with our experimental data. The only roughly matching 300 K curve is related to simulated superparamagnetic elemental

Co clusters with $d$ = 4 nm. Therefore, much steeper curves should be visible at lower temperatures. To get an almost temperature-independent behavior in the range of 0.1 T $\leq B \leq$ 1.0 T, considerably larger (*e.g.* $d$ = 20 nm) elemental Co particles have to be present. However, the initial slope of those single-domain nanoparticles has to be very steep as it is shown in Fig. S18b in this case [18–20]. For a magnetization behavior as observed in our case, almost bulk-like, or multi-domain behavior is mandatory, where demagnetizing fields help to provide less steep susceptibilities. In the case of Co metal the single domain size of perfectly spherical elemental Co particles is at least 50 nm, and it is strongly increasing for even slight variations of the spherical shape [21,22]. To obtain similar hysteresis curves as measured for our samples, elemental Co particles in the range of 80-200 nm or larger agglomerates of small Co particles are necessary. In contrast, no hints of small and large ferromagnetic Co clusters were observed in the high-resolution transmission electron microscopy (HR-TEM) investigations (Figs. 3a – e, Figs. S13–15). In addition, due to the small Co ion concentration, dipole–dipole interactions between possible present elemental Co particles are too small to explain the hysteretic behavior as observed for LTCON. Even for more than one order of magnitude higher Co ion concentration, it was not observed experimentally [18]. The combination of measured almost temperature-independent $M(H)$ curves, the calculated Langevin-plots, and the performed (HR-)TEM studies clearly exclude superparamagnetic Co clusters and large multi-domain elemental Co particles as a possible source for the HT-ferromagnetism in LTCON.

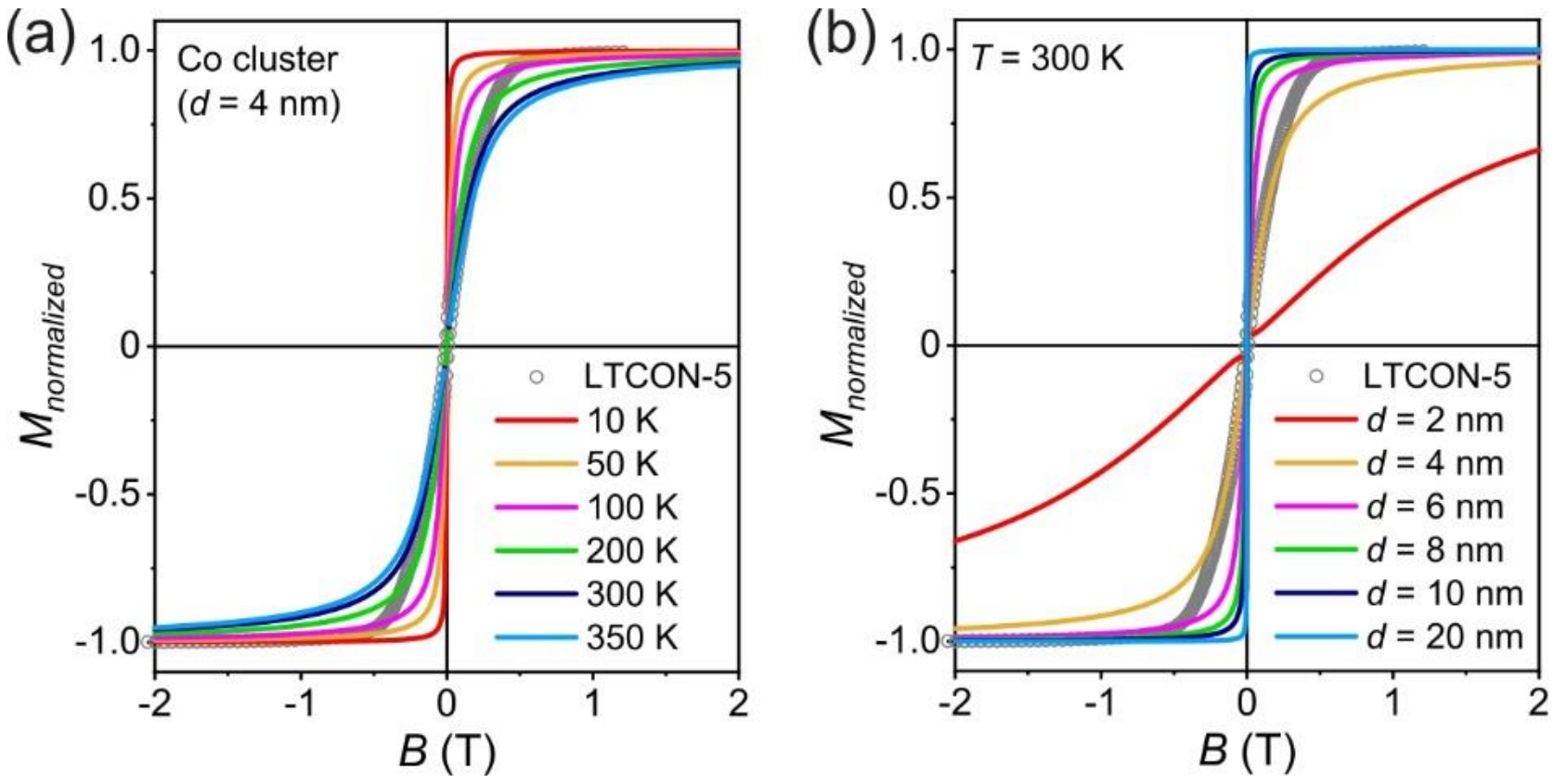

**Fig. S18.** Temperature- and size-dependent calculated $M(H)$ curves. The smallest size for possible Co clusters in the oxynitride samples is according to TEM/EDX investigations $d = 4$ nm. (a) The calculated curves of the 4 nm Co clusters were compared with the measured room temperature $M(H)$ curve at 300 K of LTCON-5. The curve shapes of the Co cluster 4 nm magnetization curves at different temperatures in comparison to those of LTCON-5 are totally different. (b) Similar comparison of the measured 300 K $M(H)$ curve at 300 K of LTCON-5 (open circles) with elemental Co cluster calculations for different cluster diameters.

**Tab. S24.** Determined coercive fields $H_c$ of LTCON. The specific Co ion concentrations $x$ are listed additionally.

| Compound | $x$ | $H_c$ (Oe) |
|---|---|---|
| **LTCON-1** | 0.01 | 188 (10 K)<br>144 (300 K) |
| **LTCON-3** | 0.03 | 243 (10 K)<br>183 (300 K) |
| **LTCON-5** | 0.05 | 245 (10 K)<br>148 (300 K) |

## Supplemental Material Sec. VI: Miscellaneous

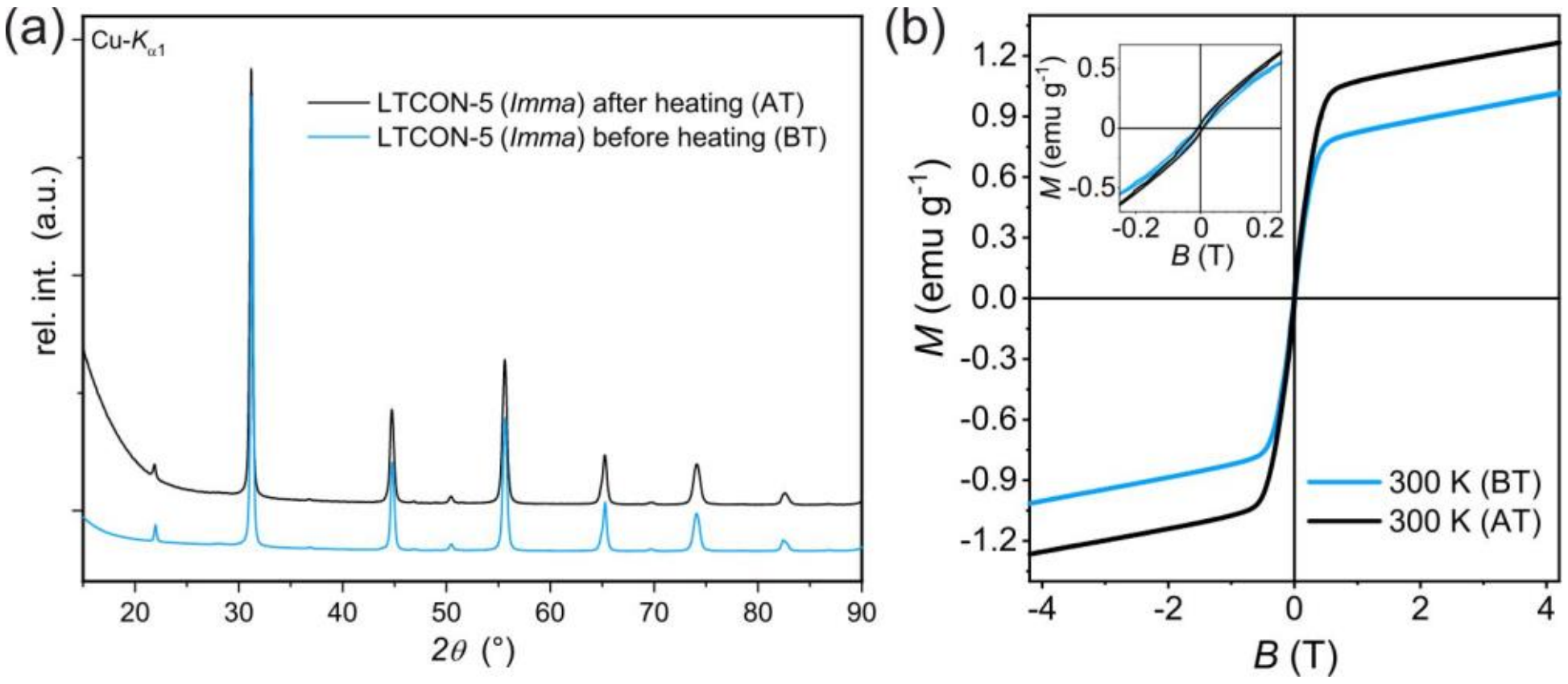

**Fig. S19.** Further high-temperature investigations of LTCON-5. (a) Complete diffractograms of LTCON-5 before the heating step (BT) and after the heating step (AT) in the SQUID furnace. (b) Hysteresis curves of LTCON-5 before the furnace measurement (BT) and afterwards (AT).

## Supplemental Material Sec. VII: LTNON

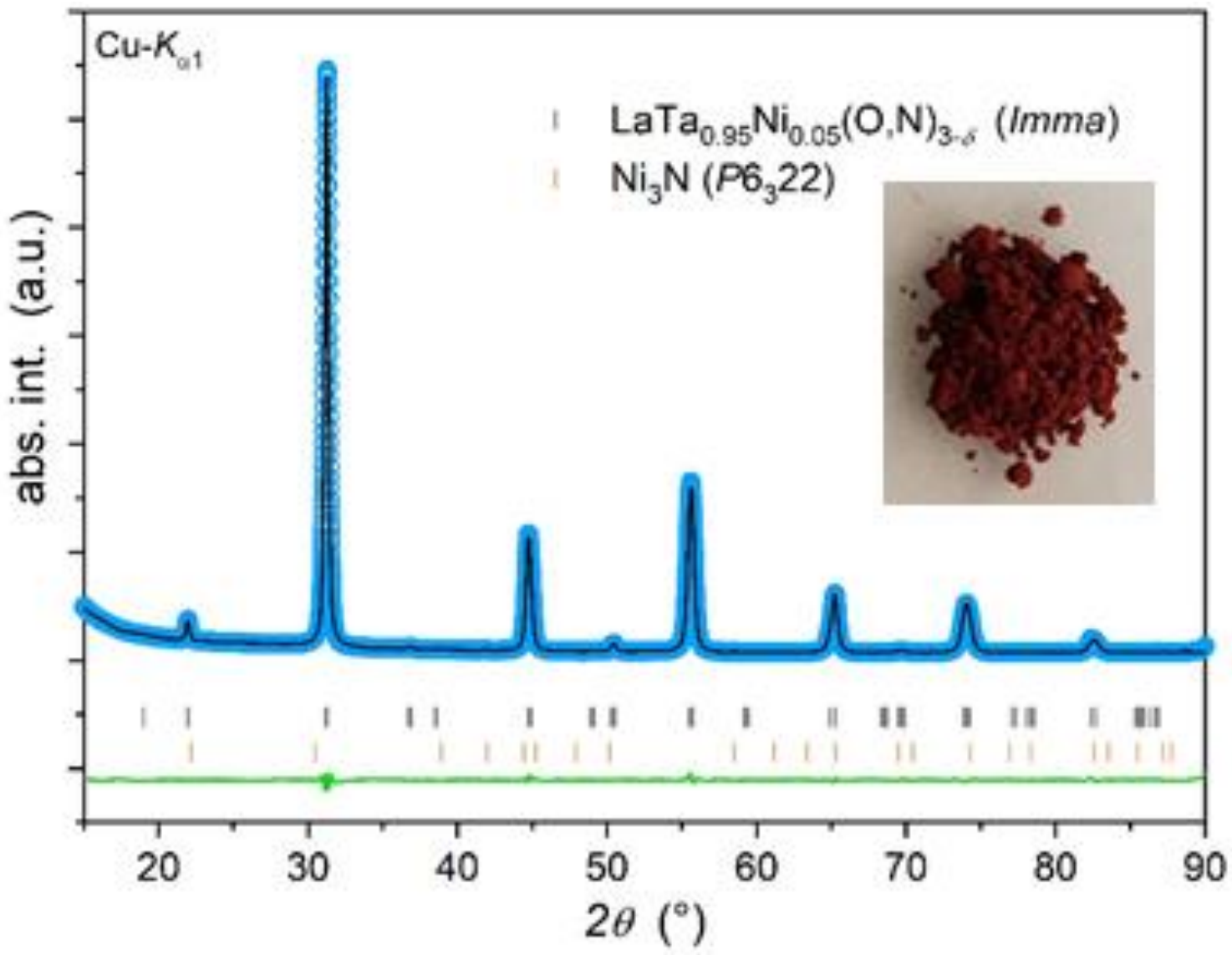

**Fig. S20.** Rietveld refinements of the PXRD data of LTNON. Secondary phase is $Ni_3N$. LTNON shows the same color as LTCON (see Foto).

**Tab. S25.** Unit cell parameters of LTNON (from PXRD data).

| $x_{Ni}$ (mol) | 0.0498 ($x_{Ni,\ expected}$ = 0.05) | |
|---|---|---|
| **Unit Cell Parameter** | $LaTa_{1-x}Ni_x(O,N)_{3-\delta}$ | $Ni_3N$ |
| **Space group type** | *Imma* | $P6_322$ |
| ***a*** **(Å)** | 5.7157(1) | 4.6271(7) |
| ***b*** **(Å)** | 8.0689(8) | 4.6271(7) |
| ***c*** **(Å)** | 5.7422(5) | 4.3079(1) |
| $\gamma$ (°) | 90 | 120 |
| $V_{cell}$ (Å³) | 264.83(2) | 79.87(3) |
| $\rho$ (g/cm³) | 8.983 | 7.904 |
| **Phase fraction (wt.-%)** | 98.1(5) | 1.8(5) |
| $R_p$ (%) | 1.92 | |
| $R_{wp}$ (%) | 2.59 | |
| $\chi^2$ | 5.53 | |
| $R_{Bragg}$ (%) | 1.20 | 8.89 |

**Tab. S26.** Refined atom positions of crystalline LTNON which revealed the maximum solubility of Ni in LTNON with $x = 0.023(2)$ (($\leq$ 0.6 at% Ni). Therefore, the real composition is determined as: $LaTa_{0.977(2)}Ni_{0.023(2)}(O,N)_3$.

| Atom | Wyck. Symb. | *x* | *y* | *z* | $B_{iso}$ (Å²) | sof[1] |
|---|---|---|---|---|---|---|
| La | 4*e* | 0 | ¼ | 0.5[4] | 1.3(6) | 1[2] |
| Ta | 4*a* | 0 | 0 | 0 | 1.1(9) | 0.977(2) |
| Ni | 4*a* | 0 | 0 | 0 | 1.1(9) | 0.023(2) |
| O(1) | 4*e* | 0 | ¼ | 0.1071(2) | 0.5[2] | 0.402[3] |
| N(1) | 4*e* | 0 | ¼ | 0.1071(2) | 0.5[2] | 0.597[3] |
| O(2) | 8*g* | ¼ | 0.9698(7) | ¼ | 0.5[2] | 0.402[3] |
| N(2) | 8*g* | ¼ | 0.9698(7) | ¼ | 0.5[2] | 0.597[3] |

[1]site occupancy factor, [2]fixed, [3]fixed according to HGE results, [4]fixed according to Porter *et al*. [11]

**Tab. S27.** Thought experiment. It shows the plausibility of the XMCD results for a 100 % ferromagnetic LTCON-5 sample.

| **Origin** | **Percent of phase** | $m_{spin}$ **($\mu_B$/$Co^{3+}$)** |
|---|---|---|
| XMCD result | 100 % ferromagnetic LTCON-5 | **1.14 ± 0.06** |
| Thought experiment | 90 % non-magnetic LTCON-5 | **0** |
| Thought experiment | 10 % ferromagnetic Co-containing phase | $10 \cdot 1.14 \sim$ **11.4** |
| Thought experiment | 90 % non-magnetic LTCON-5 + 10 % ferromagnetic Co-containing phase | $0.9 \cdot 0 + 0.1 \cdot 11.4 =$ **1.14** |

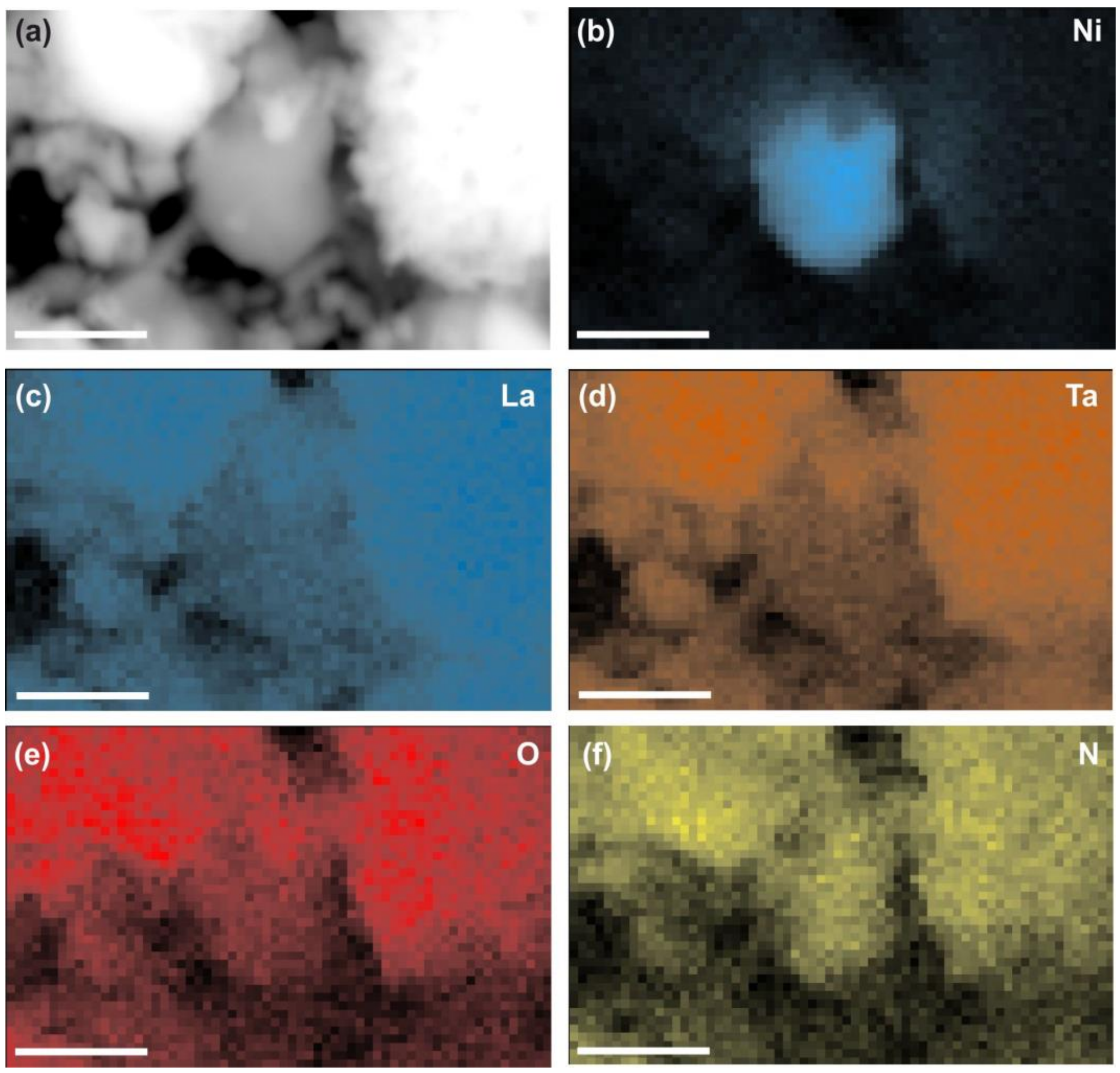

**Fig. S21.** SEM/EDX measurements of LTNON. (a) SEM image of LTNON and (b)–(f) the respective EDX measurements of Ni, La, Ta, O, and N. Scale bar: 500 nm

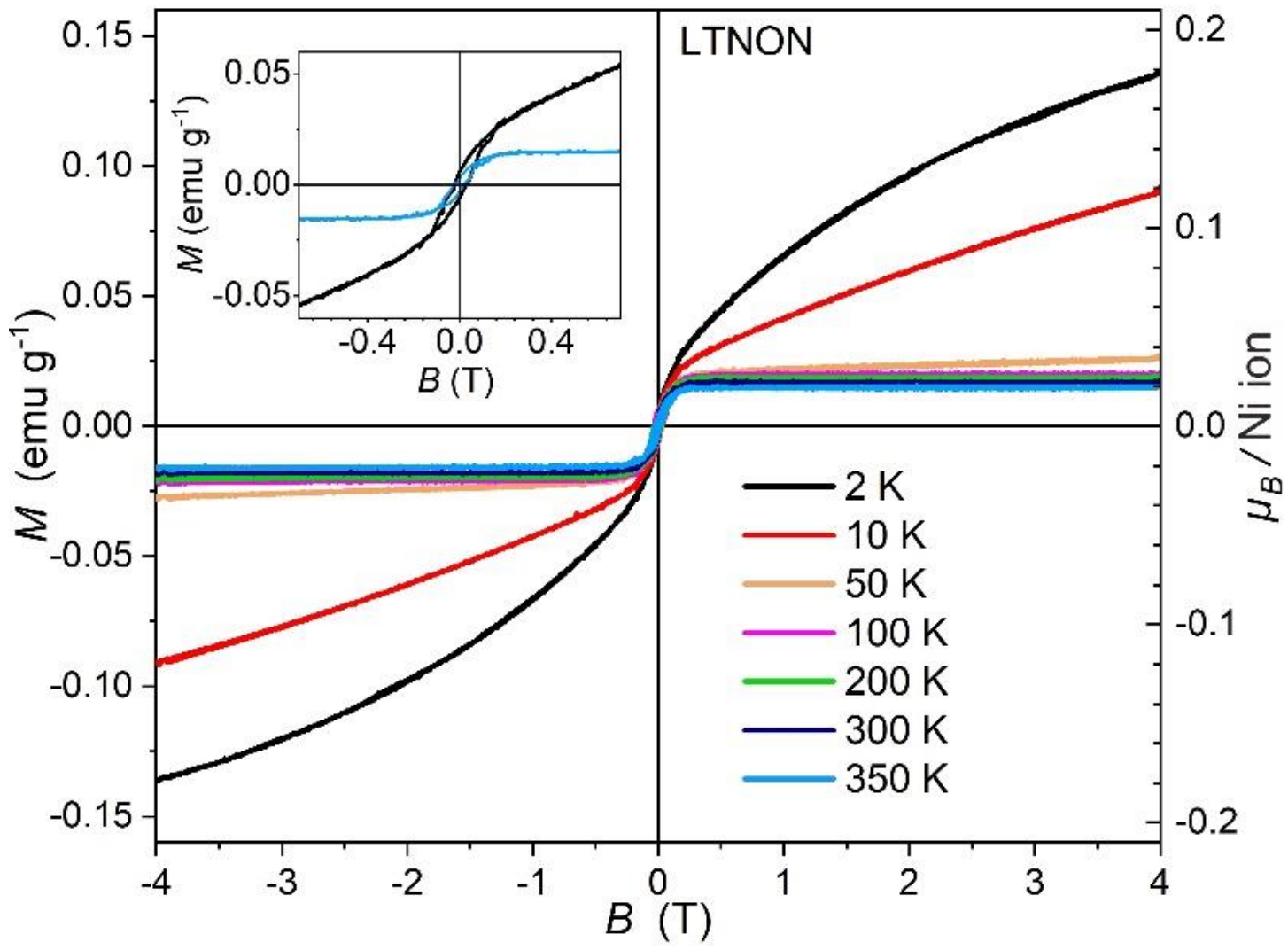

**Fig. S22.** SQUID-based magnetic measurements of LTNON.

## Supplementary References